\documentclass[aps, prl, twocolumn, superscriptaddress]{revtex4-1}

\usepackage{amsmath}
\usepackage{hyperref}
\usepackage{latexsym}
\usepackage{graphicx}
\usepackage[usenames,dvipsnames]{xcolor}
\usepackage{bm}
\usepackage{siunitx} 
\usepackage{braket} 
\usepackage[normalem]{ulem}

\newif\ifcmnt
\cmnttrue

\ifcmnt
    \providecommand{\aucmnt}[1]{#1}
\else
    \providecommand{\aucmnt}[1]{}
\fi

\begin{document}

\title{A strong loophole-free test of local realism}

\author{Lynden K. Shalm}
\affiliation{National Institute of Standards and Technology, 325 Broadway, Boulder, CO 80305, USA}

\author{Evan Meyer-Scott}
\affiliation{Institute for Quantum Computing and Department of Physics and Astronomy, University of Waterloo, 200 University Ave West, Waterloo, Ontario, Canada, N2L 3G1}

\author{Bradley G. Christensen}
\affiliation{Department of Physics, University of Illinois at Urbana-Champaign, Urbana, IL 61801, USA}

\author{Peter Bierhorst}
\affiliation{National Institute of Standards and Technology, 325 Broadway, Boulder, CO 80305, USA}

\author{Michael A. Wayne}
\affiliation{Department of Physics, University of Illinois at Urbana-Champaign, Urbana, IL 61801, USA}
\affiliation{National Institute of Standards and Technology, 100 Bureau Drive,\,Gaithersburg,\,MD 20899,\,USA}

\author{Martin J. Stevens}
\affiliation{National Institute of Standards and Technology, 325 Broadway, Boulder, CO 80305, USA}

\author{Thomas Gerrits}
\affiliation{National Institute of Standards and Technology, 325 Broadway, Boulder, CO 80305, USA}

\author{Scott Glancy}
\affiliation{National Institute of Standards and Technology, 325 Broadway, Boulder, CO 80305, USA}

\author{Deny R. Hamel}
\affiliation{D\'epartement de Physique et d'Astronomie, Universit\'e de Moncton, Moncton, New Brunswick E1A 3E9, Canada}

\author{Michael S. Allman}
\affiliation{National Institute of Standards and Technology, 325 Broadway, Boulder, CO 80305, USA}

\author{Kevin J. Coakley}
\affiliation{National Institute of Standards and Technology, 325 Broadway, Boulder, CO 80305, USA}

\author{Shellee D. Dyer}
\affiliation{National Institute of Standards and Technology, 325 Broadway, Boulder, CO 80305, USA}

\author{Carson Hodge}
\affiliation{National Institute of Standards and Technology, 325 Broadway, Boulder, CO 80305, USA}

\author{Adriana E. Lita}
\affiliation{National Institute of Standards and Technology, 325 Broadway, Boulder, CO 80305, USA}

\author{Varun B. Verma}
\affiliation{National Institute of Standards and Technology, 325 Broadway, Boulder, CO 80305, USA}

\author{Camilla Lambrocco}
\affiliation{National Institute of Standards and Technology, 325 Broadway, Boulder, CO 80305, USA}

\author{Edward Tortorici}
\affiliation{National Institute of Standards and Technology, 325 Broadway, Boulder, CO 80305, USA}

\author{Alan L. Migdall}
\affiliation{National Institute of Standards and Technology, 100 Bureau Drive,\,Gaithersburg,\,MD 20899,\,USA}
\affiliation{Joint Quantum Institute, National Institute of Standards and Technology and University of Maryland, 100 Bureau Drive, Gaithersburg, Maryland 20899, USA}

\author{Yanbao Zhang}
\affiliation{Institute for Quantum Computing and Department of Physics and Astronomy, University of Waterloo, 200 University Ave West, Waterloo, Ontario, Canada, N2L 3G1}

\author{Daniel R. Kumor}
\affiliation{Department of Physics, University of Illinois at Urbana-Champaign, Urbana, IL 61801, USA}

\author{William H. Farr}
\affiliation{Jet Propulsion Laboratory, California Institute of Technology, 4800 Oak Grove Drive, Pasadena, CA 91109}

\author{Francesco Marsili}
\affiliation{Jet Propulsion Laboratory, California Institute of Technology, 4800 Oak Grove Drive, Pasadena, CA 91109}

\author{Matthew D. Shaw}
\affiliation{Jet Propulsion Laboratory, California Institute of Technology, 4800 Oak Grove Drive, Pasadena, CA 91109}

\author{Jeffrey A. Stern}
\affiliation{Jet Propulsion Laboratory, California Institute of Technology, 4800 Oak Grove Drive, Pasadena, CA 91109}

\author{Carlos Abell\'{a}n}
\affiliation{ICFO -- Institut de Ciencies Fotoniques, The Barcelona Institute of Science and Technology, 08860 Castelldefels (Barcelona), Spain}

\author{Waldimar Amaya}
\affiliation{ICFO -- Institut de Ciencies Fotoniques, The Barcelona Institute of Science and Technology, 08860 Castelldefels (Barcelona), Spain}

\author{Valerio Pruneri}
\affiliation{ICFO -- Institut de Ciencies Fotoniques, The Barcelona Institute of Science and Technology, 08860 Castelldefels (Barcelona), Spain}
\affiliation{ICREA -- Instituci\'{o} Catalana de Recerca i Estudis Avan\c{c}ats, 08015 Barcelona, Spain}

\author{Thomas Jennewein}
\affiliation{Institute for Quantum Computing and Department of Physics and Astronomy, University of Waterloo, 200 University Ave West, Waterloo, Ontario, Canada, N2L 3G1}
\affiliation{Quantum Information Science Program, Canadian Institute for Advanced Research, Toronto, ON, Canada}

\author{Morgan W. Mitchell}
\affiliation{ICFO -- Institut de Ciencies Fotoniques, The Barcelona Institute of Science and Technology, 08860 Castelldefels (Barcelona), Spain}
\affiliation{ICREA -- Instituci\'{o} Catalana de Recerca i Estudis Avan\c{c}ats, 08015 Barcelona, Spain}

\author{Paul G. Kwiat}
\affiliation{Department of Physics, University of Illinois at Urbana-Champaign, Urbana, IL 61801, USA}

\author{Joshua C. Bienfang}
\affiliation{National Institute of Standards and Technology, 100 Bureau Drive,\,Gaithersburg,\,MD 20899,\,USA}
\affiliation{Joint Quantum Institute, National Institute of Standards and Technology and University of Maryland, 100 Bureau Drive, Gaithersburg, Maryland 20899, USA}

\author{Richard P. Mirin}
\affiliation{National Institute of Standards and Technology, 325 Broadway, Boulder, CO 80305, USA}

\author{Emanuel Knill}
\affiliation{National Institute of Standards and Technology, 325 Broadway, Boulder, CO 80305, USA}

\author{Sae Woo Nam}
\affiliation{National Institute of Standards and Technology, 325 Broadway, Boulder, CO 80305, USA}

\begin{abstract}
We present a loophole-free violation of local realism using entangled photon pairs.  We
ensure that all relevant events in our Bell test are spacelike separated by placing the
parties far enough apart and by using fast random number generators and high-speed
polarization measurements. A high-quality polarization-entangled source of photons,
combined with high-efficiency, low-noise, single-photon detectors, allows us to make
measurements without requiring any fair-sampling assumptions.  Using a hypothesis test,
we compute p-values as small as $5.9\times 10^{-9}$ for our Bell violation while
maintaining the spacelike separation of our events. We estimate the degree to which a
local realistic system could predict our measurement choices. Accounting for this
predictability, our smallest adjusted p-value is $2.3 \times 10^{-7}$. We therefore
reject the hypothesis that local realism governs our experiment.
\end{abstract}
\date{\today}
\maketitle

 \textit{But if [a hidden variable theory] is local it will not agree with quantum mechanics, and if it agrees with quantum mechanics it will not be local. This is what the theorem says.} \textsc{—--John Stewart Bell }\cite{Bell1975}
\\
\\
Quantum mechanics at its heart is a statistical theory. It cannot with certainty predict
the outcome of all single events, but instead it predicts probabilities of outcomes. This
probabilistic nature of quantum theory is at odds with the determinism inherent in
Newtonian physics and relativity, where outcomes can be exactly predicted given
sufficient knowledge of a system. Einstein and others felt that quantum mechanics was
incomplete. Perhaps quantum systems are controlled by variables, possibly hidden from us
\cite{Holland2004}, that determine the outcomes of measurements. If we had
direct access to these hidden variables, then the outcomes of all measurements performed
on quantum systems could be predicted with certainty.  De Broglie's 1927 pilot-wave
theory was a first attempt at formulating a hidden variable theory of quantum physics
\cite{Broglie1927}; it was completed in 1952 by David Bohm \cite{Bohm1952a,Bohm1952b}.
While the pilot-wave theory can reproduce all of the predictions of quantum mechanics, it
has the curious feature that hidden variables in one location can instantly change values
because of events happening in distant locations. This seemingly violates the locality
principle from relativity, which says that objects cannot signal one another faster than
the speed of light. In 1935 the nonlocal feature of quantum systems was popularized by
Einstein, Podolsky, and Rosen
\cite{Einstein1935}, and is something Einstein later referred to as ``spooky actions at a
distance''\cite{Einstein1971}. But in 1964 John Bell showed that it is impossible to
construct a hidden variable theory that obeys locality and simultaneously reproduces all
of the predictions of quantum mechanics \cite{Bell1964}. Bell's theorem fundamentally changed our
understanding of quantum theory and today stands as a cornerstone of modern quantum
information science.

Bell's theorem does not prove the validity of quantum mechanics, but it does allow us to
test the hypothesis that nature is governed by local realism. The principle of realism
says that any system has pre-existing values for all possible measurements of the
system.  In local realistic theories, these pre-existing values depend only on events in
the past lightcone of the system.  Local hidden-variable theories obey this principle of
local realism. Local realism places constraints on the behavior of systems of multiple
particles---constraints that do not apply to entangled quantum particles. This leads to
different predictions that can be tested in an experiment known as a Bell test. In a
typical two-party Bell test, a source generates particles and sends them to two distant
parties, Alice and Bob. Alice and Bob independently and randomly choose properties of
their individual particles to measure. Later, they compare the results of their
measurements. Local realism constrains the joint probability distribution of their
choices and measurements. The basis of a Bell test is an inequality that is obeyed by
local realistic probability distributions but can be violated by the probability
distributions of certain entangled quantum particles
\cite{Bell1964}. A few years after Bell derived his inequality, new forms were
introduced by Clauser, Horne, Shimony and Holt \cite{Clauser1969}, and Clauser
and Horne \cite{Clauser1974} that are simpler to experimentally test.

In a series of landmark experiments, Freedman and Clauser \cite{Freedman1972} and Aspect,
Grangier, Dalibard, and Roger \cite{Aspect1981, Aspect1982a, Aspect1982b} demonstrated
experimental violations of Bell inequalities using pairs of polarization-entangled
photons generated by an atomic cascade. However, due to technological constraints, these
Bell tests and those that followed (see \cite{Genovese2005} for a review) were forced to
make additional assumptions to show local realism was incompatible with their experimental results. A significant violation of Bell's inequality implies that either
local realism is false or that one or more of the assumptions made about the experiment
are not true; thus every assumption in an experiment opens a ``loophole.'' No experiment
can be absolutely free of all loopholes, but in \cite{Larsson2014} a minimal set of
assumptions is described that an experiment must make to be considered ``loophole free.''
Here we report a significant, loophole free, experimental violation of local realism using entangled
photon pairs. We use the definition of loophole free as
defined in \cite{Larsson2014}. In our experiment the only assumptions that remain are
those that can never---even in principle---be removed.  We present physical arguments and
evidence that these remaining assumptions are either true or untestable.

Bell's proof requires that the measurement choice at Alice cannot influence the outcome
at Bob (and vice-versa). If a signal traveling from Alice can not reach Bob in the time
between Alice's choice and the completion of Bob's measurement, then there is no way for
a local hidden variable constrained by special relativity at Alice to change Bob's
outcomes.  In this case we say that Alice and Bob are spacelike separated from one
another. If an experiment does not have this spacelike separation, then an assumption
must be made that local hidden variables cannot signal one another, leading to the ``locality'' loophole.
\begin{figure*}
\centering
\includegraphics[width=6in]{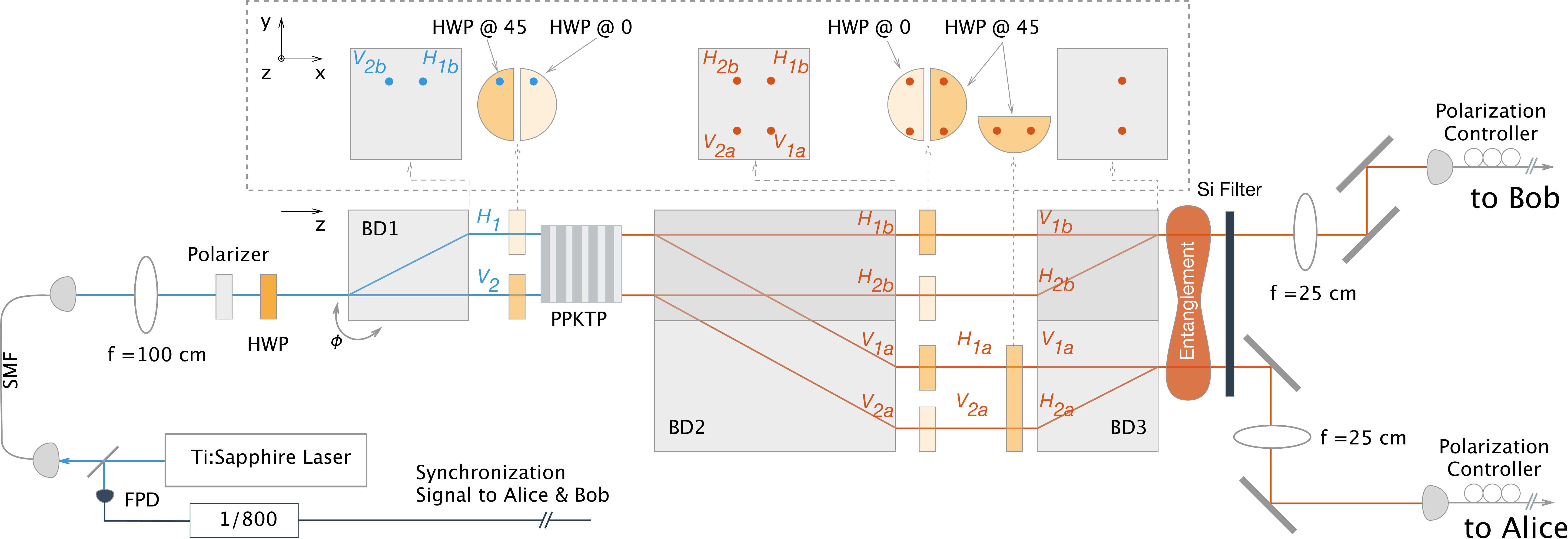}
\caption{\label{f:apparatus}
 Schematic of the entangled photon source. A pulsed \SI{775}{\nano\meter}-wavelength
 Ti:Sapphire picosecond mode-locked laser running at \SI{79.3}{\mega\hertz} repetition
 rate is used as both a clock and a pump in our setup. A fast photodiode (FPD) and
 divider circuit are used to generate the synchronization signal that is distributed to
 Alice and Bob. A polarization-maintaining single-mode fiber (SMF) then acts as a spatial
 filter for the pump. After exiting the SMF, a polarizer and half-wave plate (HWP) set
 the pump polarization. To generate entanglement, a periodically poled potassium titanyl
 phosphate (PPKTP) crystal designed for Type-II phasematching is placed in a
 polarization-based Mach-Zehnder interferometer formed using a series of HWPs and three beam displacers (BD). At BD1 the pump beam is split in two paths ($1$ and
 $2$): the horizontal (H) component of polarization of the pump translates laterally in
 the $x$ direction while the vertical (V) component of polarization passes straight
 through. Tilting BD1 sets the phase, $\phi$, of the interferometer to 0. After BD1 the
 pump state is $\left( \mathrm{cos}(16^{\circ}) \left| H_{1} \right\rangle + \mathrm{sin}(16^{\circ})
 \left| V_{2}
 \right\rangle \right)$. To address the polarization of the paths individually, semi-circular waveplates are used. A HWP in path 2 rotates the polarization of the pump from
 vertical (V) to horizontal (H). A second HWP at $0^{\circ}$ is inserted into path 1 to
 keep the path lengths of the interferometer balanced. The pump is focused at two spots
 in the crystal, and photon pairs at a wavelength of \SI{1550}{\nano\meter} are generated
 in either path 1 or 2 through the process of spontaneous parametric downconversion.
 After the crystal, BD2 walks the V-polarized signal photons down in the $y$ direction
 ($\text{V}_{1a}$ and $\text{V}_{2a}$) while the H-polarized idler photons pass straight
 through ($\text{H}_{1b}$ and $\text{H}_{2b}$). The $x$--$y$ view shows the resulting
 locations of the four beam paths. HWPs at $45^{\circ}$ correct the polarization while
 HWPs at $0^{\circ}$ provide temporal compensation. BD3 then completes the interferometer
 by recombining paths 1 and 2 for the signal and idler photons. The two downconversion
 processes interfere with one another, creating the entangled state in Eq.
 (\ref{E:BellState}). A high-purity silicon wafer with an anti-reflection coating is used
 to filter out the remaining pump light. The idler (signal) photons are coupled into a SMF and sent to Alice (Bob).
 }
\end{figure*}

\begin{figure}
\centering
\includegraphics[width=3.5in]{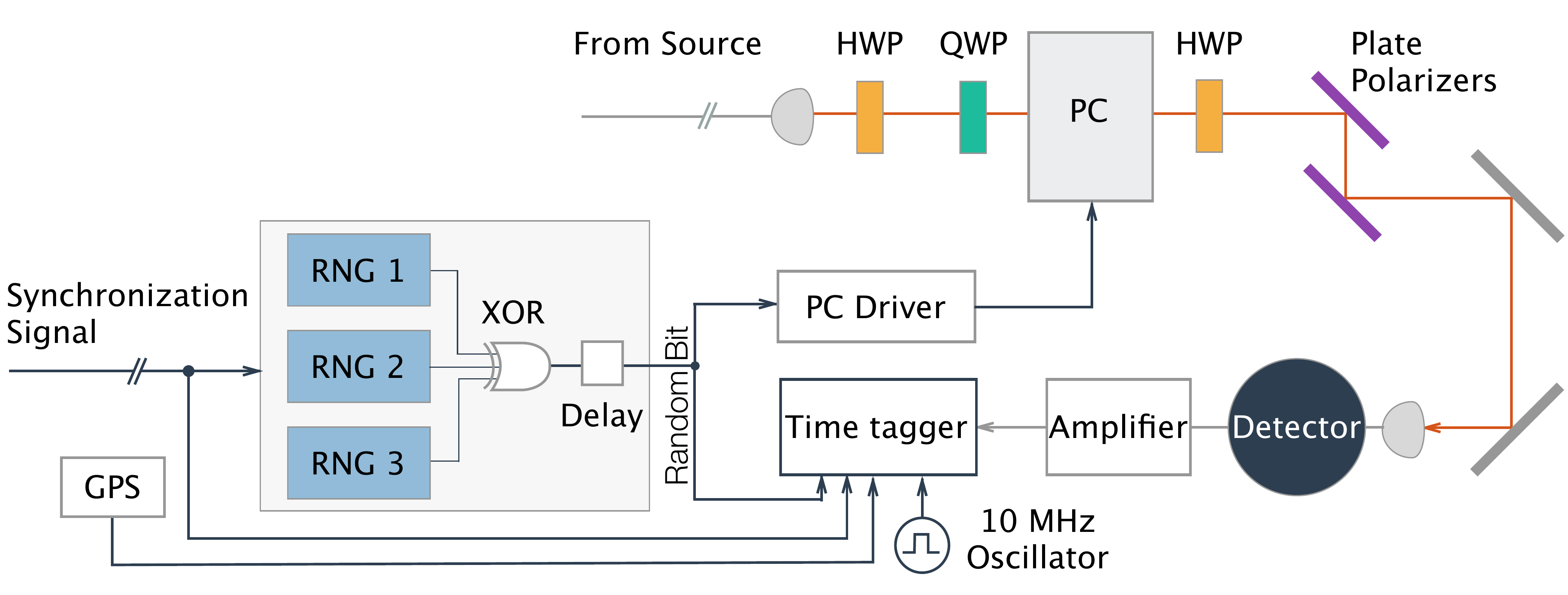}
\caption{\label{f:receiver}
 Receiver station setup for Alice and Bob. A photon arrives from the source. Two half-wave plates (HWP), a quarter-wave plate (QWP), a Pockels cell (PC), and two plate
 polarizers together act to measure the polarization state of the incoming photon. The
 polarization projection is determined by a random bit from XORing the outputs of two
 random number generators (RNG1 and RNG2) with pre-determined pseudorandom bits (RNG3).
 If the random bit is ``0'', corresponding to measurement setting  $a$ ($b$) for Alice
 (Bob), the Pockels cell remains off. If the random bit is ``1'', corresponding to
 measurement setting  $a'$ ($b'$) for Alice (Bob), then a voltage is applied to the
 Pockels cell that rotates the polarization of the photons using a fast electro-optic
 effect. The two plate polarizers have a combined contrast ratio $> 7000:1$. The photons
 are coupled back into a single-mode fiber (SMF) and detected using a superconducting
 nanowire single-photon detector (SNSPD). The signal is amplified and sent to a time-
 tagging unit where the arrival time of the event is recorded.  The time tagger also
 records the measurement setting, the synchronization signal, and a one pulse-per-second
 signal from a global positioning system (GPS). The pulse-per-second signal provides an
 external time reference that helps align the time tags Alice and Bob record. A
 \SI{10}{\mega\hertz} oscillator synchronizes the internal clocks on Alice's and Bob's
 time taggers. The synchronization pulse from the source is used to trigger the
 measurement basis choice.}
\end{figure}

Another requirement in a Bell test is that Alice and Bob must be free to make random
measurement choices that are physically independent of one another and of any properties
of the particles. If this is not true, then a hidden variable could predict the chosen
settings in advance and use that information to produce measurement outcomes that violate
a Bell inequality. Not fulfilling this requirement opens the ``freedom-of-choice''
loophole.  While this loophole can never in principle be closed, the set of hidden
variable models that are able to predict the choices can be constrained using space-like
separation.  In particular, in experiments that use processes such as cascade emission or
parametric downconversion to create entangled particles, space-like separation of the
measurement choices from the creation event eliminates the possibility that the
particles, or any other signal emanating from the creation event, influence the settings.
To satisfy this condition, Alice and Bob must choose measurement settings based on fast
random events that occur in the short time before a signal traveling at the speed of
light from the entangled-photon creation would be able to reach them. But it is
fundamentally impossible to conclusively prove that Alice's and Bob's random number
generators are independent without making additional assumptions, since their backward
lightcones necessarily intersect. Instead, it is possible to justify the assumption of
measurement independence through a detailed characterization of the physical properties
of the random number generators (such as the examination described in
\cite{Abellan2015v2arxiv,Mitchell2015}).

In any experiment, imperfections could lead to loss, and not all particles will be
detected. To violate a Bell inequality in an experiment with two parties, each free to
choose between two settings, Eberhard showed that at least 2/3 of the particles must be
detected \cite{Eberhard1993} if nonmaximally entangled states are used. If the loss
exceeds this threshold, then one may observe a violation by discarding events in which at
least one party does not detect a particle. This is valid under the assumption that
particles were lost in an unbiased manner. However, relying on this assumption opens the
``detector'' or ``fair-sampling'' loophole. While the locality and fair-sampling
loopholes have been closed individually in different systems \cite{Weihs1998,
Rowe2001,Scheidl2010, Giustina2013, Christensen2013}, it has only recently been possible to close all loopholes simultaneously using nitrogen vacancy centers in diamonds \cite{Hensen2015Nature}, and now with entangled photons in our experiment and in the work reported in \cite{Giustina2015arxiv}. These three experiments also address the freedom-of-choice loophole by space-like separation.

Fundamentally a Bell inequality is a constraint on probabilities that are estimated from
random data. Determining whether a data set shows violation is a statistical hypothesis-testing problem.  It is critical that the statistical analysis does not introduce
unnecessary assumptions that create loopholes.  A Bell test is divided into a series of
trials. In our experiment, during each trial Alice and Bob randomly choose between one
of two measurement settings (denoted $\{a, a'\}$ for Alice and $\{b, b'\}$ for Bob) and
record either a ``+'' if they observe any detection events or a ``0'' otherwise. Alice
and Bob must define when a trial is happening using only locally available information,
otherwise additional loopholes are introduced. At the end of the experiment Alice and Bob
compare the results they obtained on a trial-by-trial basis.

Our Bell test uses a version of the Clauser-Horne inequality
\cite{Clauser1974,Eberhard1993,Bierhorst2015} where, according to local
realism,
\begin{eqnarray}
\label{E:CH}
P(\text{++}\mid ab) & \leq & P(\text{+0}\mid ab')  + \notag \\ & &
 P(\text{0+}\mid a'b)+P(\text{++}\mid a'b').
\end{eqnarray}
The terms $P(\text{++}\mid ab)$ and $P(\text{++}\mid a'b')$ correspond to the probability
that both Alice and Bob record detection events (++) when they choose the measurement
settings $ab$ or $a'b'$, respectively. Similarly, the terms $P(\text{+0}\mid ab')$ and
$P(\text{0+}\mid a'b)$ are the probabilities that only Alice or Bob record an event for
settings $ab'$ and $a'b$, respectively. A local realistic model can saturate
this inequality; however, the probability distributions of entangled quantum particles
can violate it.

To quantify our Bell violation we construct a hypothesis test based on the inequality in
Eq. (\ref{E:CH}). The null hypothesis we test is that the measured probability
distributions in our experiment are constrained by local realism. Our evidence against
this null hypothesis of local realism is quantified in a p-value that we compute from our
measured data using a test statistic. Our test statistic takes all of the measured data
from Alice's and Bob's trials and summarizes them into a single number (see the
Supplemental Material for further details). The p-value is then the maximum
probability that our experiment, if it is governed by local realism, could have produced
a value of the test statistic that is at least as large as the observed value
\cite{Shao1998}. Smaller p-values can be interpreted as stronger evidence against this
hypothesis.  These p-values can also be used as certificates for cryptographic
applications, such as random number generation, that rely on a Bell test
\cite{Pironio2009,Christensen2013}.  We use a martingale binomial technique from
\cite{Bierhorst2015} for computing the p-value that makes no assumptions about the
distribution of events and does not require that the data be independent and identically
distributed \cite{Gill2003a} as long as appropriate stopping criteria are determined in
advance.

\begin{figure}
\centering
\includegraphics[width=3in]{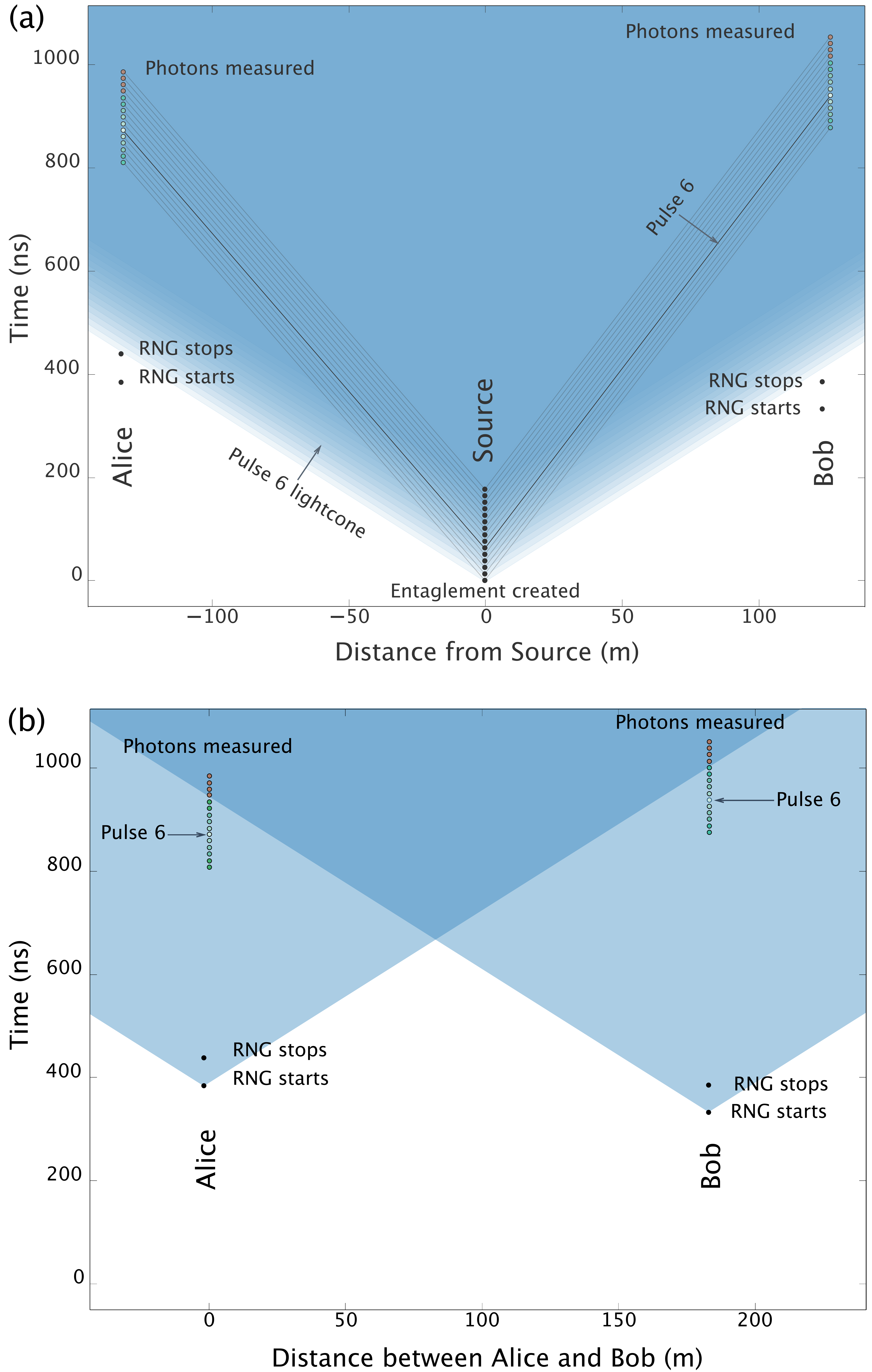}
\caption{\label{f:lightcone}
 Minkowski diagrams for the spacetime events related to Alice (A) and the source (S) and
 Bob (B) and the source (a), and Alice and Bob (b). All lightcones are shaded blue. Due
 to the geometry of Alice, Bob, and the source, more than one spacetime diagram is
 required. In a) the random number generators (RNGs) at Alice and Bob must finish picking
 a setting outside the lightcone of the birth of an entangled photon pair. A total of 15
 pump pulses have a chance of downconverting into an entangled pair of photons each time
 the Pockels cells are on. The events related to the first pulse are not spacelike
 separated, because Alice's RNG does not finish picking a setting before information
 about the properties of the photon pair can arrive; pulses 2 through 11 are spacelike
 separated. As shown in (b), pulses 12 through 15 are not spacelike separated as the
 measurement is finished by Alice and Bob after information about the other party's
 measurement setting could have arrived. In our experiment the events related to pulse 6
 are the furthest outside of all relevant lightcones. }
\end{figure}

In our experiment, the source creates polarization-entangled pairs of photons and
distributes them to Alice and Bob, located in distant labs.  At the source location a
mode-locked Ti:Sapphire laser running at repetition rate of approximately \SI{79.3}{\mega\hertz}
produces picosecond pulses centered at a wavelength of \SI{775}{\nano\meter} as shown in figure \ref{f:apparatus}. These laser pulses pump an apodized periodically poled
potassium titanyl phosphate (PPKTP) crystal to produce photon pairs at a wavelength of
\SI{1550}{\nano\meter} via the process of spontaneous parametric downconversion
\cite{Dixon2014}. The downconversion system was designed using the tools available in
\cite{spdcalc}.  The PPKTP crystal is embedded in the middle of a polarization-based
Mach-Zehnder interferometer that enables high-quality polarization-entangled states to be
generated \cite{Bennink10}. Rotating the polarization analyzer angles at Alice and Bob,
we measure the visibility of coincidence detections for a maximally entangled state to be
$0.999 \pm 0.001$ in the horizontal/vertical polarization basis and $0.996 \pm 0.001$ in
the diagonal/antidiagonal polarization basis (see \cite{uncertaintiesStatement} for information about the reported uncertainties). The entangled
photons are then coupled into separate single-mode optical fibers with one photon sent to
Alice and the other to Bob. Alice, Bob, and the source are positioned at the vertices of
a nearly right-angle triangle. Due to constraints in the building layout, the photons
travel to Alice and Bob in fiber optic cables that are not positioned along their direct
lines of sight. While the photons are in flight toward Alice and Bob, their random number
generators each choose a measurement setting. Each choice is completed before information
about the entangled state, generated at the PPKTP crystal, could possibly reach the
random number generators. When the photons arrive at Alice and Bob, they are launched
into free space, and each photon passes through a Pockels cell and polarizer that perform
the polarization measurement chosen by the random number generators as shown in Fig.
\ref{f:receiver}. After the polarizer, the photons are coupled back into a single-mode
fiber and sent to superconducting nanowire single-photon detectors, each with a detection
efficiency of $91 \pm 2\:\%$ \cite{Marsili2013}. The detector signal is then amplified
and sent to a time tagger where the arrival time is recorded. We assume the measurement
outcome is fixed when it is recorded by the time tagger, which happens before information
about the other party's setting choice could possibly arrive, as shown in Fig.
\ref{f:lightcone}(b).

\begin{figure}
\centering
\includegraphics[width=3.2in]{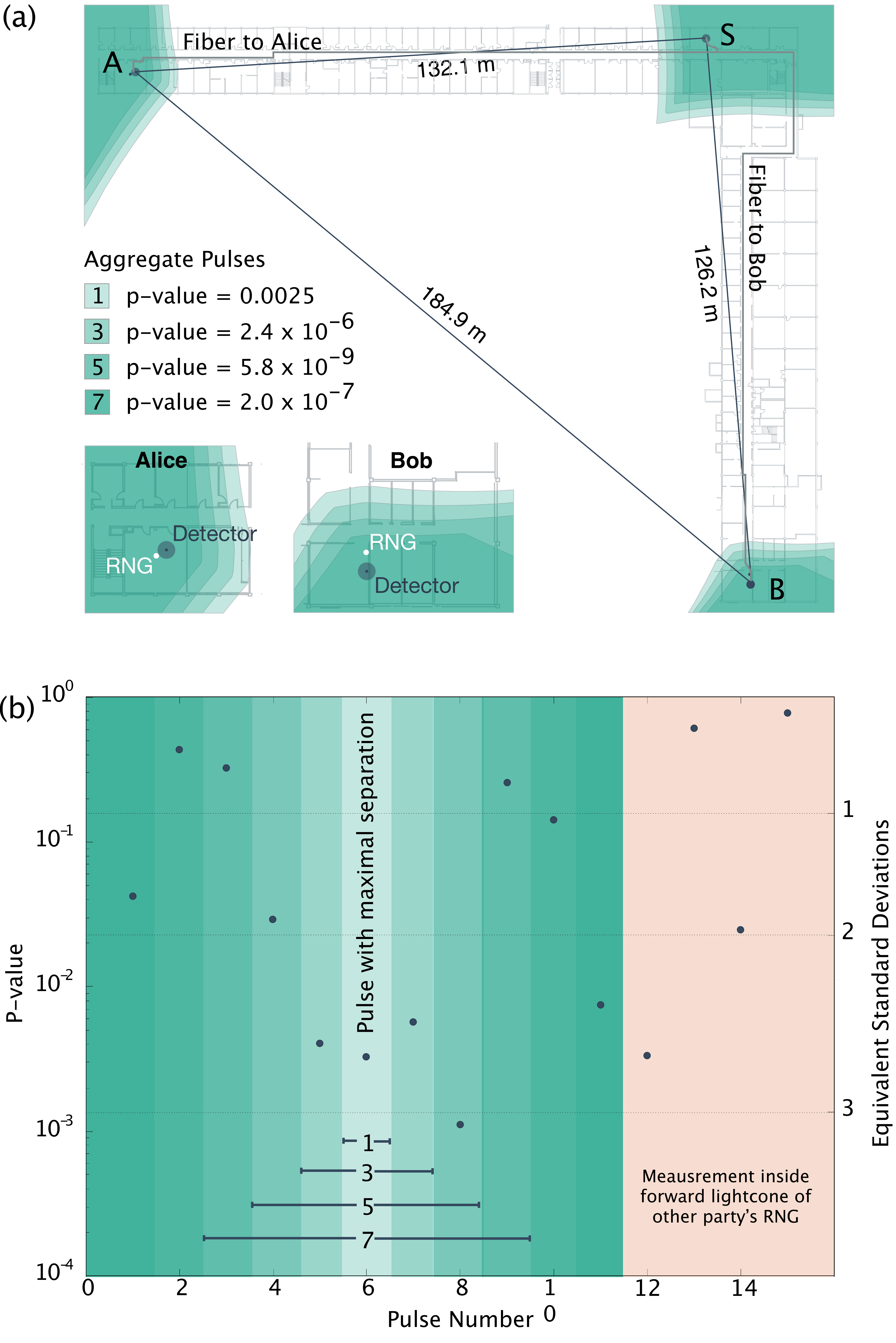}
\caption{\label{f:countours} (a) The positions of Alice (A), Bob (B), and the source (S)
  in the building where the experiment was carried out. The insets show a magnified
  ($\times 2$) view of Alice's and Bob's locations. The white dots are the location of
  the random number generators (RNGs). The larger circle at each location has a radius of
  \SI{1}{\meter} and corresponds to our uncertainty in the spatial position measurements.
  Alice, Bob, and the source can be located anywhere within the green shaded regions and
  still have their events be spacelike separated. Boundaries are plotted for aggregates
  of one, three, five, and seven pulses. Each boundary is computed by keeping the
  chronology of events fixed, but allowing the distance between the three parties to vary
  independently. In (b) the p-value of each of the individual 15 pulses is shown.
  Overlayed on the plot are the aggregate pulse combinations used in the contours in (a).
  The statistical significance of our Bell violation does not appear to depend on
  the spacelike separation of events. For reference and comparison purposes only, the
  corresponding number of standard deviations for a given p-value (for a one-sided normal
  distribution) are shown. }
\end{figure}

Alice and Bob have system detection efficiencies of $74.7 \pm 0.3\:\%$ and $75.6 \pm
0.3\:\%$, respectively. We measure this system efficiency using the method outlined by
Klyshko \cite{Klyshko80}. Background counts from blackbody radiation and room lights
reduce our observed violation of the Bell inequality. Every time a background count is
observed it counts as a detection event for only one party. These background counts
increase the heralding efficiency required to close the detector loophole above 2/3
\cite{Eberhard1993}. To reduce the number of background counts, the only detection events
considered are those that occur within a window of approximately \SI{625}{\pico\second}
at Alice and \SI{781}{\pico\second} at Bob, centered around the expected arrival times of
photons from the source. The probability of observing a background count during a single
window is $8.9\times 10^{-7}$ for Alice and $3.2\times 10^{-7}$ for Bob, while the
probability that a single pump pulse downconverts into a photon pair is $\approx 5 \times
10^{-4}$. These background counts in our system raise the efficiency needed to violate a
Bell inequality from 2/3 to 72.5\:\%. Given our system detection efficiencies, our
entangled photon production rates, entanglement visibility, and the number of background
counts, we numerically determine the entangled state and measurement settings for Alice
and Bob that should give the largest Bell violation for our setup. The optimal state is
not maximally entangled \cite{Eberhard1993} and is given by:

\begin{eqnarray}
\label{E:BellState}
\left|\psi \right\rangle = 0.961 \left|H_AH_B \right\rangle + 0.276 \left|V_AV_B \right\rangle,
\end{eqnarray}
where $H$ ($V$) denotes horizontal (vertical) polarization, and $A$
and $B$ correspond to Alice's and Bob's photons, respectively. From the simulation we also
determine that Alice's optimal polarization measurement angles, relative to a vertical
polarizer, are $\{a =4.2^o, a' = -25.9^o\}$ while Bob's are $\{b = -4.2^o, b' =
25.9^o\}$.

Synchronization signals enable Alice and Bob to define trials based only on local
information. The synchronization signal runs at a frequency of \SI{99.1}{\kilo\hertz},
allowing Alice and Bob to perform 99,100 trials/s (\SI{79.3}{\mega\hertz}/800). This
trial frequency is limited by the rate the Pockels cells can be stably driven. When the
Pockels cells are triggered they stay on for $\approx$ \SI{200}{\nano\second}. This is
more than 15 times longer than the \SI{12.6}{\nano\second} pulse-to-pulse separation of
the pump laser. Therefore photons generated by the source can arrive in one of 15 slots
while both Alice's and Bob's Pockels cells are on. Since the majority of the photon
pulses arriving in these 15 slots satisfy the spacelike separation constraints, it is
possible to aggregate multiple adjacent pulses to increase the event rate and statistical
significance of the Bell violation. However, including too many pulses will cause one or
more of the spacelike separation constraints to be violated.  Because the probability per
pulse of generating an entangled photon pair is so low, given that one photon has already arrived, the chance of getting a second event in the same Pockels cell window is negligible ($<1\:\%$).

\begin{figure} \centering \includegraphics[width=3.5in]{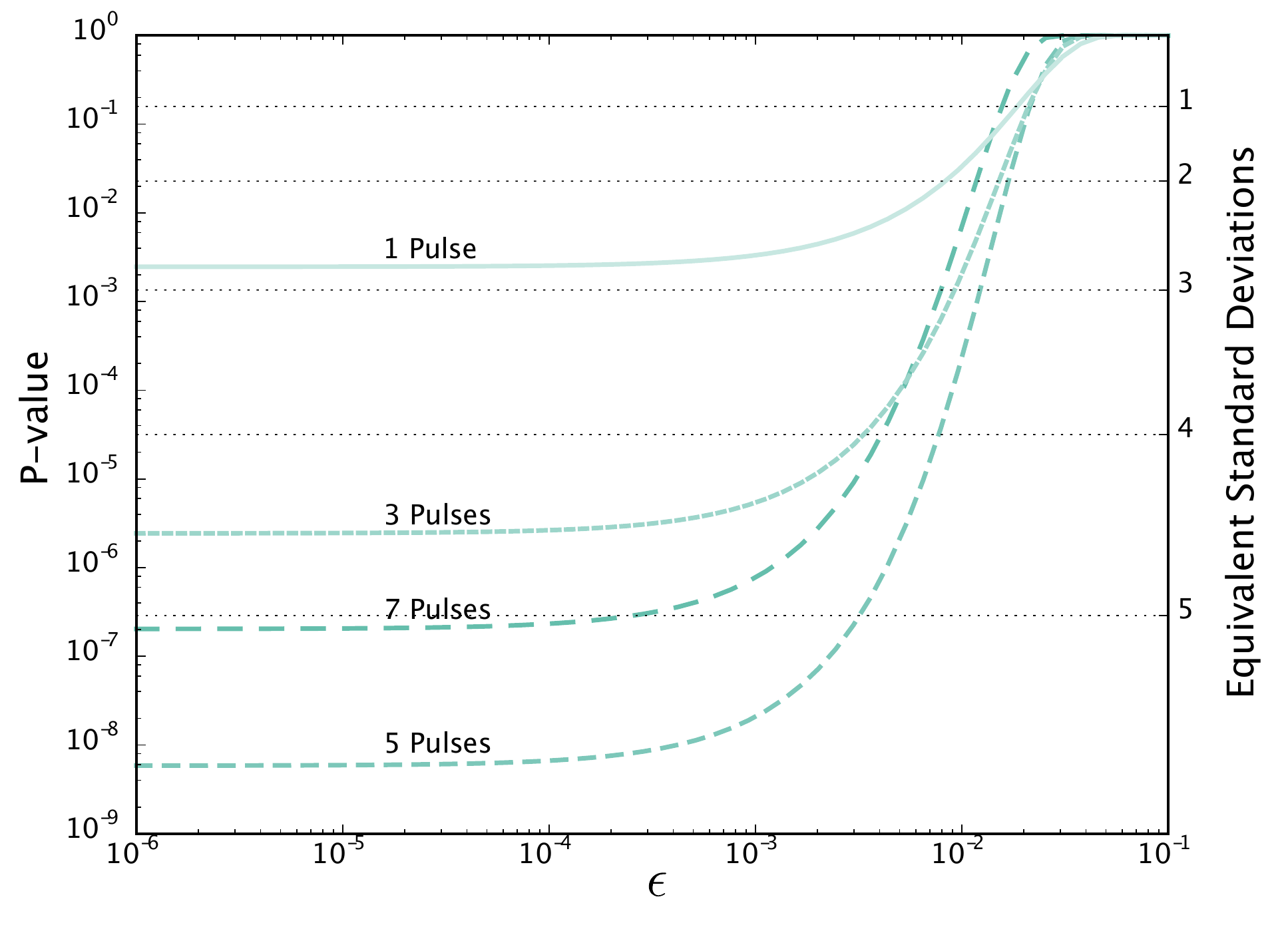}
\caption{\label{f:rngbias} The p-value for different numbers of aggregate pulses as a
function of the excess predictability, $\epsilon$, in Alice's and Bob's measurement
settings. Larger levels of predictability correspond to a weakening of the assumption
that the settings choices are physically independent of the photon properties Alice and
Bob measure. As in Fig. \ref{f:countours}(b), the p-value equivalent confidence levels
corresponding to the number of standard deviations of a one-sided normal distribution are
shown for reference. }
\end{figure}

Alice and Bob each have three different sources of random bits that they XOR together to
produce their random measurement decisions (for more information see the Supplemental
Material). The first source is based on measuring optical phase diffusion in a gain-
switched laser that is driven above and below the lasing threshold. A new bit is produced
every \SI{5} {\nano\second} by comparing adjacent laser pulses \cite{Abellan2015v2arxiv}. Each
bit is then XORed with all past bits that have been produced (for more details see the
Supplemental Material). The second source is based on sampling the amplitude of an
optical pulse at the single-photon level in a short temporal interval. This source
produces a bit on demand and is triggered by the synchronization signal. Finally, Alice
and Bob each have a different predetermined pseudorandom source that is composed of
various popular culture movies and TV shows, as well as the digits of $\pi$, XORed
together. Suppose that a local-realistic system with the goal of producing violation of
the Bell inequality, was able to manipulate the properties of the photons emitted by the
entanglement source before each trial. Provided that the randomness sources correctly
extract their bits from the underlying processes of phase diffusion, optical
amplitude sampling, and the production of cultural artifacts (such as the movie
\textit{Back to the Future}), this powerful local realistic system would be required to
predict the outcomes of all of these processes well in advance of the beginning of each
trial to achieve its goal. Such a model would have elements of superdeterminism---the
fundamentally untestable idea that all events in the universe are preordained.

Over the course of two days we took a total of 6 data runs with differing configurations
of the experimental setup \cite{onlinedata}. Here we report the results from the final
dataset that recorded data for 30 minutes (see the Supplemental Material for descriptions
and results from all datasets). This is the dataset where the experiment was most stable
and best aligned; small changes in coupling efficiency and the stability of the Pockels
cells can lead to large changes in the observed violation. The events corresponding to
the sixth pulse out of the 15 possible pulses per trial are the farthest outside all the
relevant lightcones. Thus we say these events are the most spacelike separated. To
increase our data rate we aggregate multiple pulses centered around pulse number 6. We
consider different Bell tests using a single pulse (number 6), three pulses (pulses 5, 6,
and 7), five pulses (pulses 4 through 8), and seven pulses (pulses 3 through 9). The
joint measurement outcomes and corresponding p-values for these combinations are shown in
Table \ref{t:pvalues}. For a single pulse we measure a p-value = $2.5 \times 10^{-3}$,
for three pulses a p-value = $2.4\times 10^{-6} $, for five pulses a p-value = $5.8
\times 10^{-9}$, and for seven pulses a p-value = $2.0\times 10^{-7} $, corresponding to
a strong violation of local realism.

If, trial-by-trial, a conspiratorial hidden variable (or attacker in cryptographic
scenarios) has some measure of control over or knowledge about the settings choices at
Alice and Bob, then they could manipulate the outcomes to observe a violation of a Bell
inequality.  Even if we weaken our assumption that Alice's and Bob's setting choices are
physically independent from the source, we can still compute valid p-values against the
hypothesis of local realism.  We characterize the lack of physical independence with
predictability of our random number generators.  The ``predictability,'' $\mathcal{P}$,
of a random number generator is the probability with which an adversary or local
realistic system could guess a given setting choice.  We use the parameter $\epsilon$,
the ``excess predictability'' to place an upper bound on the actual predictability of our
random number generators:
\begin{eqnarray}
\label{E:predictability}
\mathcal{P} \leq \frac{1}{2}(1+\epsilon).
\end{eqnarray}
In principle, it is impossible to measure predictability through statistical tests of the
random numbers, because they can be made to appear random, unbiased, and independent even
if the excess predictability during each trial is nonzero. Extreme examples that could
cause nonzero excess predictability include superdeterminism or a powerful and devious
adversary with access to the devices, but subtle technical issues can never be entirely
ruled out. Greater levels of excess predictability lead to lower statistical confidence
in a rejection of local realism. In Fig. \ref{f:rngbias} we show how different levels of
excess predictability change the statistical significance of our results
\cite{Bierhorst2013} (see Supplemental Material for more details). We can make
estimates of the excess predictability in our system. From additional measurements, we
observe a bias of $(1.08 \pm 0.07) \times 10^{-4}$ in the settings reaching the XOR from
the laser diffusion random source, which includes synchronization electronics as well as
the random number generator. If this bias is the only source of predictability in our
system, this level of bias would correspond to an excess predictability of approximately
$2 \times 10^{-4}$. To be conservative we use an excess predictability bound that is
fifteen times larger, $\epsilon_{p} = 3 \times 10^{-3}$ (see Supplemental Material for
more details). If our experiment had excess predictability equal to $\epsilon_{p}$ our
p-values would be increased to $5.9 \times 10^{-3}$, $2.4 \times 10^{-5}$, $2.3 \times
10^{-7}$, and $9.2 \times 10^{-6}$ for one, three, five, and seven pulses, respectively
\cite{Bierhorst2013}. Combining the output of this random number generator with the
others should lead to lower bias levels and a lower excess predictability, but even under
the the paranoid situation where a nearly superdeterministic local realistic system has
complete knowledge of the bits from the other random number sources, the adjusted
p-values still provide a rejection of local realism with high statistical significance.

\begin{table*}[]
\begin{tabular}{|c||c|c|c|c|c|c|c|}
\hline
Aggregate Pulses & $N(\text{++}\mid ab)$ & $N_{\text{stop}}$ & Total trials  & P-value              & Adjusted p-value       & Timing Margin (ns)     & Minimum distance (m)    \\ \hline
1                & 1257                  & 2376               & 175,654,992  & $2.5 \times 10^{-3}$ & $5.9 \times 10^{-3}$                 & $63.5 \pm 3.7$     &  $9.2$             \\ \hline
3                & 3800                  & 7211               & 175,744,824  & $2.4\times 10^{-6}$  & $2.4 \times 10^{-5}$   & $50.9 \pm 3.7$      &  $7.3$             \\ \hline
5                & 6378                  & 12127              & 177,358,351  & $5.9\times 10^{-9}$  & $2.3 \times 10^{-7}$   & $38.3 \pm 3.7$     &  $5.4$             \\ \hline
7                & 8820                  & 16979              & 177,797,650  & $2.0\times 10^{-7}$  & $9.2 \times 10^{-6}$   & $25.7 \pm 3.7$     &  $3.5$             \\ \hline
\end{tabular}
\caption{\label{t:pvalues} P-value results for different numbers of
  aggregate pulses. Here $N(\text{++}\mid ab)$ refers to the number of times Alice and
  Bob both detect a photon with settings $a$ and $b$ respectively. Before analyzing the
  data a stopping criteria, $N_{\text{stop}}$, was chosen. This stopping criteria refers
  to the total number of events considered that have the settings and outcomes
  specified by the terms in Eq.  (\ref{E:CH}), $N_{\text{stop}} = N(\text{++}\mid ab) +
  N(\text{+0}\mid ab') + N(\text{0+}\mid a'b) + N(\text{++}\mid a'b')$. After this number
  of trials the p-value is computed and the remaining trials discarded. Such pre-determined stopping criteria are necessary for the hypothesis test we use (see
  Supplemental Material for more details). The total trials include all trials up to the
  stopping criteria regardless of whether a photon is detected. The adjusted p-value
  accounts for the excess predictability we estimate from measurements of one of our
  random number generators. As discussed in the text, the time difference between Bob
  finishing his measurement and the earliest time at which information about Alice's measurement choice could arrive at Bob
  sets the margin of timing error that can be tolerated and still have all events
  guaranteed to be spacelike separated.  We also give the minimum distance between each
  party and its boundary line (shown in Fig. \ref{f:countours}(a)) that guarantees
  satisfaction of the spacelike separation constraints. In the Supplemental Material the
  frequencies of each combination of settings choice for 5 aggregate pulses is reported.}
\end{table*}

Satisfying the spacetime separations constraints in Fig. \ref{f:lightcone} requires
precise measurements of the locations of Alice, Bob, and the source as well as the timing
of all events. Using a combination of position measurements from a global positioning
system (GPS) receiver and site surveying, we determine the locations of Alice, Bob, and
the source with an uncertainty of $< \SI{1}{\meter}$. This uncertainty is set by the
physical size of the cryostat used to house our detectors and the uncertainty in the GPS
coordinates. There are four events that must be spacelike separated: Alice's and Bob's
measurement choice must be fixed before any signal emanating from the photon creation
event could arrive at their locations, and Alice and Bob must finish their measurements
before information from the other party's measurement choice could reach them. Due to the
slight asymmetry in the locations of Alice, Bob, and the source, the time difference
between Bob finishing his measurement and information possibly arriving about Alice's
measurement choice is always shorter than the time differences of the other three events
as shown in Fig. \ref{f:lightcone}(b). This time difference serves as a kind of margin;
our system can tolerate timing errors as large as this margin and still have all events
remain spacelike separated. For one, three, five, and seven aggregate pulses this
corresponds to a margin of $63.5 \pm 3.7$ ns, $50.9 \pm 3.7$ ns, $38.3 \pm 3.7$ ns, and
$25.7 \pm 3.7$ ns, respectively as shown in Table \ref{t:pvalues}. The uncertainty in
these timing measurements is dominated by the \SI{1}{\meter} positional uncertainty (see
Supplemental Material for further details on the timing measurements).

A way to visualize and further quantify the the spacelike separation of events is to
compute how far Alice, Bob, and the source could move from their measured position and
still be guaranteed to satisfy the locality constraints, assuming that the chronology of
all events remains fixed.  In figure \ref{f:countours}(a) Alice, Bob, and the source
locations are surrounded by shaded green regions. As long as each party remains anywhere
inside the boundaries of these regions their events are guaranteed to be spacelike
separated. There are specific configurations where all three parties can be outside the
boundaries and still be spacelike separated, but here we consider the most conspiratorial
case where all parties can collude with one another. The boundaries are overlayed on
architectural drawings of the building in which the experiment was performed.  Four
different boundaries are plotted, corresponding to the Bell test performed with one,
three, five, and seven aggregate pulses. Minimizing over the path of each boundary line,
the minimum distance that Alice, Bob, and the source are located from their respective
boundaries is \SI{9.2}{\meter}, \SI{7.3}{\meter}, \SI{5.4}{\meter}, and
\SI{3.5}{\meter} for aggregates of one pulse, three pulses, five
pulses, and seven pulses, respectively. For these pulse configurations we would have had
to place our source and detection systems physically in different rooms (or even move
outside of the building) to compromise our spacelike separation. Aggregating more than
seven pulses leads to boundaries that are less than three meters away from our measured
positions. In these cases we are not able to make strong claims about the spacelike
separation of our events.

Finally, as shown in Fig. \ref{f:countours}(b), we can compute the 15 p-values for each
of the time slots we consider that photons from the source can arrive in every trial.
Photons arriving in slots 2 through 11 are spacelike separated while photons in slots 12
through 15 are not. The photons arriving in these later slots are measured after
information from the other party's random number generator could arrive as shown in Fig.
\ref{f:lightcone}(b). It appears that spacelike separation has no discernible effect on
the statistical significance of the violation. However, we do see large slot-to-slot
fluctuation in the calculated p-values. We suspect that this is due to instability in the
applied voltage when the Pockels cell is turned on. In this case photons receive slightly
different polarization rotations depending on which slot they arrive in, leading to non-ideal measurement settings at Alice and Bob. It is because of this slot-to-slot variation
that the aggregate of seven pulses has a computed p-value larger than the five-pulse
case. Fixing this instability and using more sophisticated hypothesis test techniques
\cite{Zhang2011,Zhang2013,knill2015} will enable us to robustly increase the statistical
significance of our violation for the seven pulse case.

The experiment reported here is a commissioning run of the Bell test machine we
eventually plan to use to certify randomness. The ability to include multiple pulses in
our Bell test highlights the flexibility of our system. Our Bell test machine is capable
of high event rates, making it well suited for generating random numbers required by
cryptographic applications \cite{Pironio2009}. Future work will focus on incorporating
our Bell test machine as an additional source of real-time randomness into the National
Institute of Standards and Technology's public random number beacon
(https://beacon.nist.gov).

It has been 51 years since John Bell formulated his test of local
realism. In that time his inequality has shaped our understanding of
entanglement and quantum correlations, led to the quantum information
revolution, and transformed the study of quantum foundations.  Until
recently it has not been possible to carry out a complete and
statistically significant loophole-free Bell test. Using advances in
random number generation, photon source development, and
high-efficiency single-photon detectors, we are able to observe a
strong violation of a Bell inequality that is
loophole free, meaning that we only need to make a minimal set of
assumptions. These assumptions are that our measurements of locations
and times of events are reliable, that Alice's and Bob's measurement
outcomes are fixed at the time taggers, and that during any given
trial the random number generators at Alice and Bob are physically
independent of each other and the properties of the photons being
measured. It is impossible, even in principle, to eliminate a form of
these assumptions in any Bell test. Under these assumptions, if a
hidden variable theory is local it does not agree with our results,
and if it agrees with our results then it is not local.  \\ \\

\begin{acknowledgments}
We thank Todd Harvey for assistance with optical fiber installation, Norman Sanford for
the use of lab space, Kevin Silverman, Aephraim M. Steinberg, Rupert Ursin, Marissa
Giustina, Stephen Jordan, Dietrich Leibfried, and Paul Lett for helpful discussions, Nik
Luhrs and Kristina Meier for helping with the electronics, Andrew Novick for help with
the GPS measurements, Joseph Chapman and Malhar Jere for designing the cultural
pseudorandom numbers, and Stephen Jordan, Paul Lett, and Dietrich Leibfried for
constructive comments on the manuscript. We thank Conrad Turner Bierhorst for waiting
patiently for the computation of p-values. We dedicate this paper to the memory of our
coauthor, colleague, and friend, Jeffrey Stern. We acknowledge support for this project
provided by: DARPA (LKS, MSA, AEL, SDD, MJS, VBV, TG, RPM, SWN, WHF, FM, MDS, JAS) and
the NIST Quantum Information Program (LKS, MSA, AEL, SDD, MJS, VBV, TG, SG, PB, JCB, AM,
RPM, EK, SWN); NSF grant No. PHY 12-05870 and MURI Center for Photonic Quantum
Information Systems (ARO/ARDA Program DAAD19-03-1-0199) DARPA InPho program and the
Office of Naval Research MURI on Fundamental Research on Wavelength-Agile High-Rate
Quantum Key Distribution (QKD) in a Marine Environment, award \#N00014-13-0627 (BGC, MAW,
DRK, PGK); NSERC, CIFAR and Industry Canada (EMS, YZ, TJ); NASA (FM, MDS, WHF, JAS);
European Research Council project AQUMET, FET Proactive project QUIC, Spanish MINECO
project MAGO (Ref. FIS2011-23520) and EPEC (FIS2014-62181-EXP), Catalan 2014-SGR-1295,
the European Regional Development Fund (FEDER) grant TEC2013-46168-R, and Fundacio
Privada CELLEX (MWM, CA, WA, VP); New Brunswick Innovation Foundation (DRH).  Part of the
research was carried out at the Jet Propulsion Laboratory, California Institute of
Technology, under a contract with the National Aeronautics and Space Administration. This
work includes contributions of the National Institute of Standards and Technology, which
are not subject to U.S. copyright.
\end{acknowledgments}

\bibliography{bibLoopholeFreeBellTestcombined}

\begin{thebibliography}{41}%
\makeatletter
\providecommand \@ifxundefined [1]{%
 \@ifx{#1\undefined}
}%
\providecommand \@ifnum [1]{%
 \ifnum #1\expandafter \@firstoftwo
 \else \expandafter \@secondoftwo
 \fi
}%
\providecommand \@ifx [1]{%
 \ifx #1\expandafter \@firstoftwo
 \else \expandafter \@secondoftwo
 \fi
}%
\providecommand \natexlab [1]{#1}%
\providecommand \enquote  [1]{``#1''}%
\providecommand \bibnamefont  [1]{#1}%
\providecommand \bibfnamefont [1]{#1}%
\providecommand \citenamefont [1]{#1}%
\providecommand \href@noop [0]{\@secondoftwo}%
\providecommand \href [0]{\begingroup \@sanitize@url \@href}%
\providecommand \@href[1]{\@@startlink{#1}\@@href}%
\providecommand \@@href[1]{\endgroup#1\@@endlink}%
\providecommand \@sanitize@url [0]{\catcode `\\12\catcode `\$12\catcode
  `\&12\catcode `\#12\catcode `\^12\catcode `\_12\catcode `\%12\relax}%
\providecommand \@@startlink[1]{}%
\providecommand \@@endlink[0]{}%
\providecommand \url  [0]{\begingroup\@sanitize@url \@url }%
\providecommand \@url [1]{\endgroup\@href {#1}{\urlprefix }}%
\providecommand \urlprefix  [0]{URL }%
\providecommand \Eprint [0]{\href }%
\providecommand \doibase [0]{http://dx.doi.org/}%
\providecommand \selectlanguage [0]{\@gobble}%
\providecommand \bibinfo  [0]{\@secondoftwo}%
\providecommand \bibfield  [0]{\@secondoftwo}%
\providecommand \translation [1]{[#1]}%
\providecommand \BibitemOpen [0]{}%
\providecommand \bibitemStop [0]{}%
\providecommand \bibitemNoStop [0]{.\EOS\space}%
\providecommand \EOS [0]{\spacefactor3000\relax}%
\providecommand \BibitemShut  [1]{\csname bibitem#1\endcsname}%
\let\auto@bib@innerbib\@empty
\bibitem [{\citenamefont {Bell}(1975)}]{Bell1975}%
  \BibitemOpen
  \bibfield  {author} {\bibinfo {author} {\bibfnamefont {J.~S.}\ \bibnamefont
  {Bell}},\ }\href@noop {} {\bibfield  {journal} {\bibinfo  {journal}
  {Epistemological Letters}\ ,\ \bibinfo {pages} {2}} (\bibinfo {year}
  {1975})}\BibitemShut {NoStop}%
\bibitem [{\citenamefont {Holland}(2005)}]{Holland2004}%
  \BibitemOpen
  \bibfield  {author} {\bibinfo {author} {\bibfnamefont {P.}~\bibnamefont
  {Holland}},\ }\href {\doibase 10.1007/s10701-004-1940-7} {\bibfield
  {journal} {\bibinfo  {journal} {Found. Phys.}\ }\textbf {\bibinfo {volume}
  {35}},\ \bibinfo {pages} {177} (\bibinfo {year} {2005})},\ \Eprint
  {http://arxiv.org/abs/quant-ph/0401017} {arXiv:quant-ph/0401017} \BibitemShut
  {NoStop}%
\bibitem [{\citenamefont {de~Broglie}(1927)}]{Broglie1927}%
  \BibitemOpen
  \bibfield  {author} {\bibinfo {author} {\bibfnamefont {L.}~\bibnamefont
  {de~Broglie}},\ }\href {\doibase 10.1051/jphysrad:0192700805022500}
  {\bibfield  {journal} {\bibinfo  {journal} {J. Phys. Radium}\ }\textbf
  {\bibinfo {volume} {8}},\ \bibinfo {pages} {225} (\bibinfo {year}
  {1927})}\BibitemShut {NoStop}%
\bibitem [{\citenamefont {Bohm}(1952{\natexlab{a}})}]{Bohm1952a}%
  \BibitemOpen
  \bibfield  {author} {\bibinfo {author} {\bibfnamefont {D.}~\bibnamefont
  {Bohm}},\ }\href {\doibase 10.1103/PhysRev.85.166} {\bibfield  {journal}
  {\bibinfo  {journal} {Phys. Rev.}\ }\textbf {\bibinfo {volume} {85}},\
  \bibinfo {pages} {166} (\bibinfo {year} {1952}{\natexlab{a}})}\BibitemShut
  {NoStop}%
\bibitem [{\citenamefont {Bohm}(1952{\natexlab{b}})}]{Bohm1952b}%
  \BibitemOpen
  \bibfield  {author} {\bibinfo {author} {\bibfnamefont {D.}~\bibnamefont
  {Bohm}},\ }\href {\doibase 10.1103/PhysRev.85.180} {\bibfield  {journal}
  {\bibinfo  {journal} {Phys. Rev.}\ }\textbf {\bibinfo {volume} {85}},\
  \bibinfo {pages} {180} (\bibinfo {year} {1952}{\natexlab{b}})}\BibitemShut
  {NoStop}%
\bibitem [{\citenamefont {Einstein}\ \emph {et~al.}(1935)\citenamefont
  {Einstein}, \citenamefont {Podolosky},\ and\ \citenamefont
  {Rosen}}]{Einstein1935}%
  \BibitemOpen
  \bibfield  {author} {\bibinfo {author} {\bibfnamefont {A.}~\bibnamefont
  {Einstein}}, \bibinfo {author} {\bibfnamefont {B.}~\bibnamefont {Podolosky}},
  \ and\ \bibinfo {author} {\bibfnamefont {N.}~\bibnamefont {Rosen}},\ }\href
  {\doibase 10.1103/PhysRev.47.777} {\bibfield  {journal} {\bibinfo  {journal}
  {Phys. Rev.}\ }\textbf {\bibinfo {volume} {47}},\ \bibinfo {pages} {777}
  (\bibinfo {year} {1935})}\BibitemShut {NoStop}%
\bibitem [{\citenamefont {Einstein}\ \emph {et~al.}(1971)\citenamefont
  {Einstein}, \citenamefont {Born},\ and\ \citenamefont {Born}}]{Einstein1971}%
  \BibitemOpen
  \bibfield  {author} {\bibinfo {author} {\bibfnamefont {A.}~\bibnamefont
  {Einstein}}, \bibinfo {author} {\bibfnamefont {M.}~\bibnamefont {Born}}, \
  and\ \bibinfo {author} {\bibfnamefont {H.}~\bibnamefont {Born}},\ }\href@noop
  {} {\emph {\bibinfo {title} {{The Born-Einstein Letters: the Correspondence
  between Max {\&} Hedwig Born and Albert Einstein 1916/1955}}}},\ \bibinfo
  {edition} {1st}\ ed.\ (\bibinfo  {publisher} {The MacMillan Press Ltd},\
  \bibinfo {address} {London and Basingstoke},\ \bibinfo {year}
  {1971})\BibitemShut {NoStop}%
\bibitem [{\citenamefont {Bell}(1964)}]{Bell1964}%
  \BibitemOpen
  \bibfield  {author} {\bibinfo {author} {\bibfnamefont {J.~S.}\ \bibnamefont
  {Bell}},\ }\href@noop {} {\bibfield  {journal} {\bibinfo  {journal}
  {Physics}\ }\textbf {\bibinfo {volume} {1}},\ \bibinfo {pages} {195}
  (\bibinfo {year} {1964})}\BibitemShut {NoStop}%
\bibitem [{\citenamefont {Clauser}\ \emph {et~al.}(1969)\citenamefont
  {Clauser}, \citenamefont {Horne}, \citenamefont {Shimony},\ and\
  \citenamefont {Holt}}]{Clauser1969}%
  \BibitemOpen
  \bibfield  {author} {\bibinfo {author} {\bibfnamefont {J.~F.}\ \bibnamefont
  {Clauser}}, \bibinfo {author} {\bibfnamefont {M.~A.}\ \bibnamefont {Horne}},
  \bibinfo {author} {\bibfnamefont {A.}~\bibnamefont {Shimony}}, \ and\
  \bibinfo {author} {\bibfnamefont {R.~A.}\ \bibnamefont {Holt}},\ }\href
  {\doibase 10.1103/PhysRevLett.23.880} {\bibfield  {journal} {\bibinfo
  {journal} {Phys. Rev. Lett.}\ }\textbf {\bibinfo {volume} {23}},\ \bibinfo
  {pages} {880} (\bibinfo {year} {1969})}\BibitemShut {NoStop}%
\bibitem [{\citenamefont {Clauser}\ and\ \citenamefont
  {Horne}(1974)}]{Clauser1974}%
  \BibitemOpen
  \bibfield  {author} {\bibinfo {author} {\bibfnamefont {J.~F.}\ \bibnamefont
  {Clauser}}\ and\ \bibinfo {author} {\bibfnamefont {M.~A.}\ \bibnamefont
  {Horne}},\ }\href@noop {} {\bibfield  {journal} {\bibinfo  {journal} {Phys.
  Rev. D}\ }\textbf {\bibinfo {volume} {10}},\ \bibinfo {pages} {526} (\bibinfo
  {year} {1974})}\BibitemShut {NoStop}%
\bibitem [{\citenamefont {Freedman}\ and\ \citenamefont
  {Clauser}(1972)}]{Freedman1972}%
  \BibitemOpen
  \bibfield  {author} {\bibinfo {author} {\bibfnamefont {S.~J.}\ \bibnamefont
  {Freedman}}\ and\ \bibinfo {author} {\bibfnamefont {J.~F.}\ \bibnamefont
  {Clauser}},\ }\href {\doibase 10.1103/PhysRevLett.28.938} {\bibfield
  {journal} {\bibinfo  {journal} {Phys. Rev. Lett.}\ }\textbf {\bibinfo
  {volume} {28}},\ \bibinfo {pages} {938} (\bibinfo {year} {1972})}\BibitemShut
  {NoStop}%
\bibitem [{\citenamefont {Aspect}\ \emph {et~al.}(1981)\citenamefont {Aspect},
  \citenamefont {Grangier},\ and\ \citenamefont {Roger}}]{Aspect1981}%
  \BibitemOpen
  \bibfield  {author} {\bibinfo {author} {\bibfnamefont {A.}~\bibnamefont
  {Aspect}}, \bibinfo {author} {\bibfnamefont {P.}~\bibnamefont {Grangier}}, \
  and\ \bibinfo {author} {\bibfnamefont {G.}~\bibnamefont {Roger}},\ }\href
  {\doibase 10.1103/PhysRevLett.47.460} {\bibfield  {journal} {\bibinfo
  {journal} {Phys. Rev. Lett.}\ }\textbf {\bibinfo {volume} {47}},\ \bibinfo
  {pages} {460} (\bibinfo {year} {1981})}\BibitemShut {NoStop}%
\bibitem [{\citenamefont {Aspect}\ \emph
  {et~al.}(1982{\natexlab{a}})\citenamefont {Aspect}, \citenamefont
  {Grangier},\ and\ \citenamefont {Roger}}]{Aspect1982a}%
  \BibitemOpen
  \bibfield  {author} {\bibinfo {author} {\bibfnamefont {A.}~\bibnamefont
  {Aspect}}, \bibinfo {author} {\bibfnamefont {P.}~\bibnamefont {Grangier}}, \
  and\ \bibinfo {author} {\bibfnamefont {G.}~\bibnamefont {Roger}},\ }\href
  {\doibase 10.1103/PhysRevLett.49.91} {\bibfield  {journal} {\bibinfo
  {journal} {Phys. Rev. Lett.}\ }\textbf {\bibinfo {volume} {49}},\ \bibinfo
  {pages} {91} (\bibinfo {year} {1982}{\natexlab{a}})}\BibitemShut {NoStop}%
\bibitem [{\citenamefont {Aspect}\ \emph
  {et~al.}(1982{\natexlab{b}})\citenamefont {Aspect}, \citenamefont
  {Dalibard},\ and\ \citenamefont {Roger}}]{Aspect1982b}%
  \BibitemOpen
  \bibfield  {author} {\bibinfo {author} {\bibfnamefont {A.}~\bibnamefont
  {Aspect}}, \bibinfo {author} {\bibfnamefont {J.}~\bibnamefont {Dalibard}}, \
  and\ \bibinfo {author} {\bibfnamefont {G.}~\bibnamefont {Roger}},\ }\href
  {\doibase 10.1103/PhysRevLett.49.1804} {\bibfield  {journal} {\bibinfo
  {journal} {Phys. Rev. Lett.}\ }\textbf {\bibinfo {volume} {49}},\ \bibinfo
  {pages} {1804} (\bibinfo {year} {1982}{\natexlab{b}})}\BibitemShut {NoStop}%
\bibitem [{\citenamefont {Genovese}(2005)}]{Genovese2005}%
  \BibitemOpen
  \bibfield  {author} {\bibinfo {author} {\bibfnamefont {M.}~\bibnamefont
  {Genovese}},\ }\href {\doibase
  http://dx.doi.org/10.1016/j.physrep.2005.03.003} {\bibfield  {journal}
  {\bibinfo  {journal} {Phys. Rep.}\ }\textbf {\bibinfo {volume} {413}},\
  \bibinfo {pages} {319 } (\bibinfo {year} {2005})},\ \Eprint
  {http://arxiv.org/abs/quant-ph/0701071} {arXiv:quant-ph/0701071} \BibitemShut
  {NoStop}%
\bibitem [{\citenamefont {Larsson}(2014)}]{Larsson2014}%
  \BibitemOpen
  \bibfield  {author} {\bibinfo {author} {\bibfnamefont {J.-{\r{A}}.}\
  \bibnamefont {Larsson}},\ }\href
  {http://stacks.iop.org/1751-8121/47/i=42/a=424003} {\bibfield  {journal}
  {\bibinfo  {journal} {J. Phys. A}\ }\textbf {\bibinfo {volume} {47}},\
  \bibinfo {pages} {424003} (\bibinfo {year} {2014})},\ \Eprint
  {http://arxiv.org/abs/1407.0363} {arXiv:1407.0363} \BibitemShut {NoStop}%
\bibitem [{\citenamefont {{Abellan}}\ \emph {et~al.}(2015)\citenamefont
  {{Abellan}}, \citenamefont {{Amaya}}, \citenamefont {{Mitrani}},
  \citenamefont {{Pruneri}},\ and\ \citenamefont
  {{Mitchell}}}]{Abellan2015v2arxiv}%
  \BibitemOpen
  \bibfield  {author} {\bibinfo {author} {\bibfnamefont {C.}~\bibnamefont
  {{Abellan}}}, \bibinfo {author} {\bibfnamefont {W.}~\bibnamefont {{Amaya}}},
  \bibinfo {author} {\bibfnamefont {D.}~\bibnamefont {{Mitrani}}}, \bibinfo
  {author} {\bibfnamefont {V.}~\bibnamefont {{Pruneri}}}, \ and\ \bibinfo
  {author} {\bibfnamefont {M.~W.}\ \bibnamefont {{Mitchell}}},\ }\href@noop {}
  {\bibfield  {journal} {\bibinfo  {journal} {ArXiv e-prints}\ } (\bibinfo
  {year} {2015})},\ \Eprint {http://arxiv.org/abs/1506.02712} {arXiv:1506.02712
  [quant-ph]} \BibitemShut {NoStop}%
\bibitem [{\citenamefont {Mitchell}\ \emph {et~al.}(2015)\citenamefont
  {Mitchell}, \citenamefont {Abellan},\ and\ \citenamefont
  {Amaya}}]{Mitchell2015}%
  \BibitemOpen
  \bibfield  {author} {\bibinfo {author} {\bibfnamefont {M.~W.}\ \bibnamefont
  {Mitchell}}, \bibinfo {author} {\bibfnamefont {C.}~\bibnamefont {Abellan}}, \
  and\ \bibinfo {author} {\bibfnamefont {W.}~\bibnamefont {Amaya}},\ }\href
  {\doibase 10.1103/PhysRevA.91.012314} {\bibfield  {journal} {\bibinfo
  {journal} {Phys. Rev. A}\ }\textbf {\bibinfo {volume} {91}},\ \bibinfo
  {pages} {012314} (\bibinfo {year} {2015})}\BibitemShut {NoStop}%
\bibitem [{\citenamefont {Eberhard}(1993)}]{Eberhard1993}%
  \BibitemOpen
  \bibfield  {author} {\bibinfo {author} {\bibfnamefont {P.~H.}\ \bibnamefont
  {Eberhard}},\ }\href {\doibase 10.1103/PhysRevA.47.R747} {\bibfield
  {journal} {\bibinfo  {journal} {Phys. Rev. A}\ }\textbf {\bibinfo {volume}
  {47}},\ \bibinfo {pages} {R747} (\bibinfo {year} {1993})}\BibitemShut
  {NoStop}%
\bibitem [{\citenamefont {Weihs}\ \emph {et~al.}(1998)\citenamefont {Weihs},
  \citenamefont {Jennewein}, \citenamefont {Simon}, \citenamefont
  {Weinfurter},\ and\ \citenamefont {Zeilinger}}]{Weihs1998}%
  \BibitemOpen
  \bibfield  {author} {\bibinfo {author} {\bibfnamefont {G.}~\bibnamefont
  {Weihs}}, \bibinfo {author} {\bibfnamefont {T.}~\bibnamefont {Jennewein}},
  \bibinfo {author} {\bibfnamefont {C.}~\bibnamefont {Simon}}, \bibinfo
  {author} {\bibfnamefont {H.}~\bibnamefont {Weinfurter}}, \ and\ \bibinfo
  {author} {\bibfnamefont {A.}~\bibnamefont {Zeilinger}},\ }\href {\doibase
  10.1103/PhysRevLett.81.5039} {\bibfield  {journal} {\bibinfo  {journal}
  {Phys. Rev. Lett.}\ }\textbf {\bibinfo {volume} {81}},\ \bibinfo {pages}
  {5039} (\bibinfo {year} {1998})},\ \Eprint {http://arxiv.org/abs/9810080}
  {arXiv:9810080} \BibitemShut {NoStop}%
\bibitem [{\citenamefont {Rowe}\ \emph {et~al.}(2001)\citenamefont {Rowe},
  \citenamefont {Kielpinski}, \citenamefont {Meyer}, \citenamefont {Sackett},
  \citenamefont {Itano}, \citenamefont {Monroe},\ and\ \citenamefont
  {Wineland}}]{Rowe2001}%
  \BibitemOpen
  \bibfield  {author} {\bibinfo {author} {\bibfnamefont {M.~A.}\ \bibnamefont
  {Rowe}}, \bibinfo {author} {\bibfnamefont {D.}~\bibnamefont {Kielpinski}},
  \bibinfo {author} {\bibfnamefont {V.}~\bibnamefont {Meyer}}, \bibinfo
  {author} {\bibfnamefont {C.~A.}\ \bibnamefont {Sackett}}, \bibinfo {author}
  {\bibfnamefont {W.~M.}\ \bibnamefont {Itano}}, \bibinfo {author}
  {\bibfnamefont {C.}~\bibnamefont {Monroe}}, \ and\ \bibinfo {author}
  {\bibfnamefont {D.~J.}\ \bibnamefont {Wineland}},\ }\href {\doibase
  10.1038/35057215} {\bibfield  {journal} {\bibinfo  {journal} {Nature}\
  }\textbf {\bibinfo {volume} {409}},\ \bibinfo {pages} {791} (\bibinfo {year}
  {2001})}\BibitemShut {NoStop}%
\bibitem [{\citenamefont {Scheidl}\ \emph {et~al.}(2010)\citenamefont
  {Scheidl}, \citenamefont {Ursin}, \citenamefont {Kofler}, \citenamefont
  {Ramelow}, \citenamefont {Ma}, \citenamefont {Herbst}, \citenamefont
  {Ratschbacher}, \citenamefont {Fedrizzi}, \citenamefont {Langford},
  \citenamefont {Jennewein},\ and\ \citenamefont {Zeilinger}}]{Scheidl2010}%
  \BibitemOpen
  \bibfield  {author} {\bibinfo {author} {\bibfnamefont {T.}~\bibnamefont
  {Scheidl}}, \bibinfo {author} {\bibfnamefont {R.}~\bibnamefont {Ursin}},
  \bibinfo {author} {\bibfnamefont {J.}~\bibnamefont {Kofler}}, \bibinfo
  {author} {\bibfnamefont {S.}~\bibnamefont {Ramelow}}, \bibinfo {author}
  {\bibfnamefont {X.-S.}\ \bibnamefont {Ma}}, \bibinfo {author} {\bibfnamefont
  {T.}~\bibnamefont {Herbst}}, \bibinfo {author} {\bibfnamefont
  {L.}~\bibnamefont {Ratschbacher}}, \bibinfo {author} {\bibfnamefont
  {A.}~\bibnamefont {Fedrizzi}}, \bibinfo {author} {\bibfnamefont {N.~K.}\
  \bibnamefont {Langford}}, \bibinfo {author} {\bibfnamefont {T.}~\bibnamefont
  {Jennewein}}, \ and\ \bibinfo {author} {\bibfnamefont {A.}~\bibnamefont
  {Zeilinger}},\ }\href {\doibase 10.1073/pnas.1002780107} {\bibfield
  {journal} {\bibinfo  {journal} {Proc. Nat. Acad. Sci. USA}\ }\textbf
  {\bibinfo {volume} {107}},\ \bibinfo {pages} {19708} (\bibinfo {year}
  {2010})},\ \Eprint {http://arxiv.org/abs/0811.3129} {arXiv:0811.3129}
  \BibitemShut {NoStop}%
\bibitem [{\citenamefont {Giustina}\ \emph {et~al.}(2013)\citenamefont
  {Giustina}, \citenamefont {Mech}, \citenamefont {Ramelow}, \citenamefont
  {Wittmann}, \citenamefont {Kofler}, \citenamefont {Beyer}, \citenamefont
  {Lita}, \citenamefont {Calkins}, \citenamefont {Gerrits}, \citenamefont
  {Nam}, \citenamefont {Ursin},\ and\ \citenamefont
  {Zeilinger}}]{Giustina2013}%
  \BibitemOpen
  \bibfield  {author} {\bibinfo {author} {\bibfnamefont {M.}~\bibnamefont
  {Giustina}}, \bibinfo {author} {\bibfnamefont {A.}~\bibnamefont {Mech}},
  \bibinfo {author} {\bibfnamefont {S.}~\bibnamefont {Ramelow}}, \bibinfo
  {author} {\bibfnamefont {B.}~\bibnamefont {Wittmann}}, \bibinfo {author}
  {\bibfnamefont {J.}~\bibnamefont {Kofler}}, \bibinfo {author} {\bibfnamefont
  {J.}~\bibnamefont {Beyer}}, \bibinfo {author} {\bibfnamefont
  {A.}~\bibnamefont {Lita}}, \bibinfo {author} {\bibfnamefont {B.}~\bibnamefont
  {Calkins}}, \bibinfo {author} {\bibfnamefont {T.}~\bibnamefont {Gerrits}},
  \bibinfo {author} {\bibfnamefont {S.~W.}\ \bibnamefont {Nam}}, \bibinfo
  {author} {\bibfnamefont {R.}~\bibnamefont {Ursin}}, \ and\ \bibinfo {author}
  {\bibfnamefont {A.}~\bibnamefont {Zeilinger}},\ }\href {\doibase
  10.1038/nature12012} {\bibfield  {journal} {\bibinfo  {journal} {Nature}\
  }\textbf {\bibinfo {volume} {497}},\ \bibinfo {pages} {227} (\bibinfo {year}
  {2013})},\ \Eprint {http://arxiv.org/abs/1212.0533} {arXiv:1212.0533}
  \BibitemShut {NoStop}%
\bibitem [{\citenamefont {Christensen}\ \emph {et~al.}(2013)\citenamefont
  {Christensen}, \citenamefont {McCusker}, \citenamefont {Altepeter},
  \citenamefont {Calkins}, \citenamefont {Gerrits}, \citenamefont {Lita},
  \citenamefont {Miller}, \citenamefont {Shalm}, \citenamefont {Zhang},
  \citenamefont {Nam}, \citenamefont {Brunner}, \citenamefont {Lim},
  \citenamefont {Gisin},\ and\ \citenamefont {Kwiat}}]{Christensen2013}%
  \BibitemOpen
  \bibfield  {author} {\bibinfo {author} {\bibfnamefont {B.~G.}\ \bibnamefont
  {Christensen}}, \bibinfo {author} {\bibfnamefont {K.~T.}\ \bibnamefont
  {McCusker}}, \bibinfo {author} {\bibfnamefont {J.~B.}\ \bibnamefont
  {Altepeter}}, \bibinfo {author} {\bibfnamefont {B.}~\bibnamefont {Calkins}},
  \bibinfo {author} {\bibfnamefont {T.}~\bibnamefont {Gerrits}}, \bibinfo
  {author} {\bibfnamefont {A.~E.}\ \bibnamefont {Lita}}, \bibinfo {author}
  {\bibfnamefont {A.}~\bibnamefont {Miller}}, \bibinfo {author} {\bibfnamefont
  {L.~K.}\ \bibnamefont {Shalm}}, \bibinfo {author} {\bibfnamefont
  {Y.}~\bibnamefont {Zhang}}, \bibinfo {author} {\bibfnamefont {S.~W.}\
  \bibnamefont {Nam}}, \bibinfo {author} {\bibfnamefont {N.}~\bibnamefont
  {Brunner}}, \bibinfo {author} {\bibfnamefont {C.~C.~W.}\ \bibnamefont {Lim}},
  \bibinfo {author} {\bibfnamefont {N.}~\bibnamefont {Gisin}}, \ and\ \bibinfo
  {author} {\bibfnamefont {P.~G.}\ \bibnamefont {Kwiat}},\ }\href {\doibase
  10.1103/PhysRevLett.111.130406} {\bibfield  {journal} {\bibinfo  {journal}
  {Phys. Rev. Lett.}\ }\textbf {\bibinfo {volume} {111}},\ \bibinfo {pages}
  {130406} (\bibinfo {year} {2013})},\ \Eprint {http://arxiv.org/abs/1306.5772}
  {arXiv:1306.5772} \BibitemShut {NoStop}%
\bibitem [{\citenamefont {Hensen}\ \emph {et~al.}(2015)\citenamefont {Hensen},
  \citenamefont {Bernien}, \citenamefont {Dreau}, \citenamefont {Reiserer},
  \citenamefont {Kalb}, \citenamefont {Blok}, \citenamefont {Ruitenberg},
  \citenamefont {Vermeulen}, \citenamefont {Schouten}, \citenamefont {Abellan},
  \citenamefont {Amaya}, \citenamefont {Pruneri}, \citenamefont {Mitchell},
  \citenamefont {Markham}, \citenamefont {Twitchen}, \citenamefont {Elkouss},
  \citenamefont {Wehner}, \citenamefont {Taminiau},\ and\ \citenamefont
  {Hanson}}]{Hensen2015Nature}%
  \BibitemOpen
  \bibfield  {author} {\bibinfo {author} {\bibfnamefont {B.}~\bibnamefont
  {Hensen}}, \bibinfo {author} {\bibfnamefont {H.}~\bibnamefont {Bernien}},
  \bibinfo {author} {\bibfnamefont {A.~E.}\ \bibnamefont {Dreau}}, \bibinfo
  {author} {\bibfnamefont {A.}~\bibnamefont {Reiserer}}, \bibinfo {author}
  {\bibfnamefont {N.}~\bibnamefont {Kalb}}, \bibinfo {author} {\bibfnamefont
  {M.~S.}\ \bibnamefont {Blok}}, \bibinfo {author} {\bibfnamefont
  {J.}~\bibnamefont {Ruitenberg}}, \bibinfo {author} {\bibfnamefont {R.~F.~L.}\
  \bibnamefont {Vermeulen}}, \bibinfo {author} {\bibfnamefont {R.~N.}\
  \bibnamefont {Schouten}}, \bibinfo {author} {\bibfnamefont {C.}~\bibnamefont
  {Abellan}}, \bibinfo {author} {\bibfnamefont {W.}~\bibnamefont {Amaya}},
  \bibinfo {author} {\bibfnamefont {V.}~\bibnamefont {Pruneri}}, \bibinfo
  {author} {\bibfnamefont {M.~W.}\ \bibnamefont {Mitchell}}, \bibinfo {author}
  {\bibfnamefont {M.}~\bibnamefont {Markham}}, \bibinfo {author} {\bibfnamefont
  {D.~J.}\ \bibnamefont {Twitchen}}, \bibinfo {author} {\bibfnamefont
  {D.}~\bibnamefont {Elkouss}}, \bibinfo {author} {\bibfnamefont
  {S.}~\bibnamefont {Wehner}}, \bibinfo {author} {\bibfnamefont {T.~H.}\
  \bibnamefont {Taminiau}}, \ and\ \bibinfo {author} {\bibfnamefont
  {R.}~\bibnamefont {Hanson}},\ }\href {http://dx.doi.org/10.1038/nature15759}
  {\bibfield  {journal} {\bibinfo  {journal} {Nature}\ }\textbf {\bibinfo
  {volume} {526}},\ \bibinfo {pages} {682} (\bibinfo {year}
  {2015})}\BibitemShut {NoStop}%
\bibitem [{\citenamefont {{Giustina}}\ \emph {et~al.}(2015)\citenamefont
  {{Giustina}}, \citenamefont {{Versteegh}}, \citenamefont {{Wengerowsky}},
  \citenamefont {{Handsteiner}}, \citenamefont {{Hochrainer}}, \citenamefont
  {{Phelan}}, \citenamefont {{Steinlechner}}, \citenamefont {{Kofler}},
  \citenamefont {{Larsson}}, \citenamefont {{Abellan}}, \citenamefont
  {{Amaya}}, \citenamefont {{Pruneri}}, \citenamefont {{Mitchell}},
  \citenamefont {{Beyer}}, \citenamefont {{Gerrits}}, \citenamefont {{Lita}},
  \citenamefont {{Shalm}}, \citenamefont {{Nam}}, \citenamefont {{Scheidl}},
  \citenamefont {{Ursin}}, \citenamefont {{Wittmann}},\ and\ \citenamefont
  {{Zeilinger}}}]{Giustina2015arxiv}%
  \BibitemOpen
  \bibfield  {author} {\bibinfo {author} {\bibfnamefont {M.}~\bibnamefont
  {{Giustina}}}, \bibinfo {author} {\bibfnamefont {M.~A.~M.}\ \bibnamefont
  {{Versteegh}}}, \bibinfo {author} {\bibfnamefont {S.}~\bibnamefont
  {{Wengerowsky}}}, \bibinfo {author} {\bibfnamefont {J.}~\bibnamefont
  {{Handsteiner}}}, \bibinfo {author} {\bibfnamefont {A.}~\bibnamefont
  {{Hochrainer}}}, \bibinfo {author} {\bibfnamefont {K.}~\bibnamefont
  {{Phelan}}}, \bibinfo {author} {\bibfnamefont {F.}~\bibnamefont
  {{Steinlechner}}}, \bibinfo {author} {\bibfnamefont {J.}~\bibnamefont
  {{Kofler}}}, \bibinfo {author} {\bibfnamefont {J.-A.}\ \bibnamefont
  {{Larsson}}}, \bibinfo {author} {\bibfnamefont {C.}~\bibnamefont
  {{Abellan}}}, \bibinfo {author} {\bibfnamefont {W.}~\bibnamefont {{Amaya}}},
  \bibinfo {author} {\bibfnamefont {V.}~\bibnamefont {{Pruneri}}}, \bibinfo
  {author} {\bibfnamefont {M.~W.}\ \bibnamefont {{Mitchell}}}, \bibinfo
  {author} {\bibfnamefont {J.}~\bibnamefont {{Beyer}}}, \bibinfo {author}
  {\bibfnamefont {T.}~\bibnamefont {{Gerrits}}}, \bibinfo {author}
  {\bibfnamefont {A.~E.}\ \bibnamefont {{Lita}}}, \bibinfo {author}
  {\bibfnamefont {L.~K.}\ \bibnamefont {{Shalm}}}, \bibinfo {author}
  {\bibfnamefont {S.~W.}\ \bibnamefont {{Nam}}}, \bibinfo {author}
  {\bibfnamefont {T.}~\bibnamefont {{Scheidl}}}, \bibinfo {author}
  {\bibfnamefont {R.}~\bibnamefont {{Ursin}}}, \bibinfo {author} {\bibfnamefont
  {B.}~\bibnamefont {{Wittmann}}}, \ and\ \bibinfo {author} {\bibfnamefont
  {A.}~\bibnamefont {{Zeilinger}}},\ }\href@noop {} {\bibfield  {journal}
  {\bibinfo  {journal} {ArXiv e-prints}\ } (\bibinfo {year} {2015})},\ \Eprint
  {http://arxiv.org/abs/1511.03190} {arXiv:1511.03190 [quant-ph]} \BibitemShut
  {NoStop}%
\bibitem [{\citenamefont {Bierhorst}(2015)}]{Bierhorst2015}%
  \BibitemOpen
  \bibfield  {author} {\bibinfo {author} {\bibfnamefont {P.}~\bibnamefont
  {Bierhorst}},\ }\href {http://stacks.iop.org/1751-8121/48/i=19/a=195302}
  {\bibfield  {journal} {\bibinfo  {journal} {J. Phys. A}\ }\textbf {\bibinfo
  {volume} {48}},\ \bibinfo {pages} {195302} (\bibinfo {year}
  {2015})}\BibitemShut {NoStop}%
\bibitem [{\citenamefont {Shao}(1998)}]{Shao1998}%
  \BibitemOpen
  \bibfield  {author} {\bibinfo {author} {\bibfnamefont {J.}~\bibnamefont
  {Shao}},\ }\href@noop {} {\emph {\bibinfo {title} {Mathematical
  Statistics}}},\ Springer Texts in Statistics\ (\bibinfo  {publisher}
  {Springer},\ \bibinfo {address} {New York},\ \bibinfo {year} {1998})\
  \bibinfo {note} {{See} 2nd edition pages 126-127.}\BibitemShut {Stop}%
\bibitem [{\citenamefont {Pironio}\ \emph {et~al.}(2010)\citenamefont
  {Pironio}, \citenamefont {Ac{\'\i}n}, \citenamefont {Massar}, \citenamefont
  {de~la Giroday}, \citenamefont {Matsukevich}, \citenamefont {Maunz},
  \citenamefont {Olmschenk}, \citenamefont {Hayes}, \citenamefont {Luo},
  \citenamefont {Manning},\ and\ \citenamefont {Monroe}}]{Pironio2009}%
  \BibitemOpen
  \bibfield  {author} {\bibinfo {author} {\bibfnamefont {S.}~\bibnamefont
  {Pironio}}, \bibinfo {author} {\bibfnamefont {A.}~\bibnamefont {Ac{\'\i}n}},
  \bibinfo {author} {\bibfnamefont {S.}~\bibnamefont {Massar}}, \bibinfo
  {author} {\bibfnamefont {A.~B.}\ \bibnamefont {de~la Giroday}}, \bibinfo
  {author} {\bibfnamefont {D.~N.}\ \bibnamefont {Matsukevich}}, \bibinfo
  {author} {\bibfnamefont {P.}~\bibnamefont {Maunz}}, \bibinfo {author}
  {\bibfnamefont {S.}~\bibnamefont {Olmschenk}}, \bibinfo {author}
  {\bibfnamefont {D.}~\bibnamefont {Hayes}}, \bibinfo {author} {\bibfnamefont
  {L.}~\bibnamefont {Luo}}, \bibinfo {author} {\bibfnamefont {T.~A.}\
  \bibnamefont {Manning}}, \ and\ \bibinfo {author} {\bibfnamefont
  {C.}~\bibnamefont {Monroe}},\ }\href@noop {} {\bibfield  {journal} {\bibinfo
  {journal} {Nature}\ }\textbf {\bibinfo {volume} {464}},\ \bibinfo {pages}
  {1021} (\bibinfo {year} {2010})},\ \Eprint {http://arxiv.org/abs/0911.3427}
  {arXiv:0911.3427} \BibitemShut {NoStop}%
\bibitem [{\citenamefont {Gill}(2003)}]{Gill2003a}%
  \BibitemOpen
  \bibfield  {author} {\bibinfo {author} {\bibfnamefont {R.~D.}\ \bibnamefont
  {Gill}},\ }in\ \href@noop {} {\emph {\bibinfo {booktitle} {Mathematical
  Statistics and Applications: Festschrift for Constance van Eeden}}},\
  Vol.~\bibinfo {volume} {42},\ \bibinfo {editor} {edited by\ \bibinfo {editor}
  {\bibfnamefont {M.}~\bibnamefont {Moore}}, \bibinfo {editor} {\bibfnamefont
  {S.}~\bibnamefont {Froda}}, \ and\ \bibinfo {editor} {\bibfnamefont
  {C.}~\bibnamefont {L{\'e}ger}}}\ (\bibinfo  {publisher} {Institute of
  Mathematical Statistics. Beachwood, Ohio},\ \bibinfo {year} {2003})\ pp.\
  \bibinfo {pages} {133--154},\ \Eprint {http://arxiv.org/abs/quant-ph/0110137}
  {arXiv:quant-ph/0110137} \BibitemShut {NoStop}%
\bibitem [{\citenamefont {Dixon}\ \emph {et~al.}(2014)\citenamefont {Dixon},
  \citenamefont {Rosenberg}, \citenamefont {Stelmakh}, \citenamefont {Grein},
  \citenamefont {Bennink}, \citenamefont {Dauler}, \citenamefont {Kerman},
  \citenamefont {Molnar},\ and\ \citenamefont {Wong}}]{Dixon2014}%
  \BibitemOpen
  \bibfield  {author} {\bibinfo {author} {\bibfnamefont {P.~B.}\ \bibnamefont
  {Dixon}}, \bibinfo {author} {\bibfnamefont {D.}~\bibnamefont {Rosenberg}},
  \bibinfo {author} {\bibfnamefont {V.}~\bibnamefont {Stelmakh}}, \bibinfo
  {author} {\bibfnamefont {M.~E.}\ \bibnamefont {Grein}}, \bibinfo {author}
  {\bibfnamefont {R.~S.}\ \bibnamefont {Bennink}}, \bibinfo {author}
  {\bibfnamefont {E.~A.}\ \bibnamefont {Dauler}}, \bibinfo {author}
  {\bibfnamefont {A.~J.}\ \bibnamefont {Kerman}}, \bibinfo {author}
  {\bibfnamefont {R.~J.}\ \bibnamefont {Molnar}}, \ and\ \bibinfo {author}
  {\bibfnamefont {F.~N.~C.}\ \bibnamefont {Wong}},\ }\href {\doibase
  10.1103/PhysRevA.90.043804} {\bibfield  {journal} {\bibinfo  {journal} {Phys.
  Rev. A}\ }\textbf {\bibinfo {volume} {90}},\ \bibinfo {pages} {043804}
  (\bibinfo {year} {2014})}\BibitemShut {NoStop}%
\bibitem [{\citenamefont {Shalm}\ \emph {et~al.}()\citenamefont {Shalm},
  \citenamefont {Garay}, \citenamefont {Palfree}, \citenamefont {Migdall},
  \citenamefont {U'Ren},\ and\ \citenamefont {Nam}}]{spdcalc}%
  \BibitemOpen
  \bibfield  {author} {\bibinfo {author} {\bibfnamefont {L.~K.}\ \bibnamefont
  {Shalm}}, \bibinfo {author} {\bibfnamefont {K.}~\bibnamefont {Garay}},
  \bibinfo {author} {\bibfnamefont {J.}~\bibnamefont {Palfree}}, \bibinfo
  {author} {\bibfnamefont {A.~L.}\ \bibnamefont {Migdall}}, \bibinfo {author}
  {\bibfnamefont {A.}~\bibnamefont {U'Ren}}, \ and\ \bibinfo {author}
  {\bibfnamefont {S.~W.}\ \bibnamefont {Nam}},\ }\href {http://www.spdcalc.org}
  {\enquote {\bibinfo {title} {Spontaneous parametric downcoversion
  calculator},}\ }\bibinfo {howpublished} {http://www.spdcalc.org}\BibitemShut
  {NoStop}%
\bibitem [{\citenamefont {Evans}\ \emph {et~al.}(2010)\citenamefont {Evans},
  \citenamefont {Bennink}, \citenamefont {Grice}, \citenamefont {Humble},\ and\
  \citenamefont {Schaake}}]{Bennink10}%
  \BibitemOpen
  \bibfield  {author} {\bibinfo {author} {\bibfnamefont {P.~G.}\ \bibnamefont
  {Evans}}, \bibinfo {author} {\bibfnamefont {R.~S.}\ \bibnamefont {Bennink}},
  \bibinfo {author} {\bibfnamefont {W.~P.}\ \bibnamefont {Grice}}, \bibinfo
  {author} {\bibfnamefont {T.~S.}\ \bibnamefont {Humble}}, \ and\ \bibinfo
  {author} {\bibfnamefont {J.}~\bibnamefont {Schaake}},\ }\href {\doibase
  10.1103/PhysRevLett.105.253601} {\bibfield  {journal} {\bibinfo  {journal}
  {Phys. Rev. Lett.}\ }\textbf {\bibinfo {volume} {105}},\ \bibinfo {pages}
  {253601} (\bibinfo {year} {2010})}\BibitemShut {NoStop}%
\bibitem [{unc()}]{uncertaintiesStatement}%
  \BibitemOpen
  \href@noop {} {}\bibinfo {howpublished} {All uncertainties $U$ and error bars
  correspond to an estimated standard deviation, $\sigma$, and a coverage
  factor $k=1$ as $U=k\sigma$}\BibitemShut {NoStop}%
\bibitem [{\citenamefont {Marsili}\ \emph {et~al.}(2013)\citenamefont
  {Marsili}, \citenamefont {Verma}, \citenamefont {Stern}, \citenamefont
  {Harrington}, \citenamefont {Lita}, \citenamefont {Gerrits}, \citenamefont
  {Vayshenker}, \citenamefont {Baek}, \citenamefont {Shaw}, \citenamefont
  {Mirin},\ and\ \citenamefont {Nam}}]{Marsili2013}%
  \BibitemOpen
  \bibfield  {author} {\bibinfo {author} {\bibfnamefont {F.}~\bibnamefont
  {Marsili}}, \bibinfo {author} {\bibfnamefont {V.~B.}\ \bibnamefont {Verma}},
  \bibinfo {author} {\bibfnamefont {J.~A.}\ \bibnamefont {Stern}}, \bibinfo
  {author} {\bibfnamefont {S.}~\bibnamefont {Harrington}}, \bibinfo {author}
  {\bibfnamefont {A.~E.}\ \bibnamefont {Lita}}, \bibinfo {author}
  {\bibfnamefont {T.}~\bibnamefont {Gerrits}}, \bibinfo {author} {\bibfnamefont
  {I.}~\bibnamefont {Vayshenker}}, \bibinfo {author} {\bibfnamefont
  {B.}~\bibnamefont {Baek}}, \bibinfo {author} {\bibfnamefont {M.~D.}\
  \bibnamefont {Shaw}}, \bibinfo {author} {\bibfnamefont {R.~P.}\ \bibnamefont
  {Mirin}}, \ and\ \bibinfo {author} {\bibfnamefont {S.~W.}\ \bibnamefont
  {Nam}},\ }\href {\doibase 10.1038/nphoton.2013.13} {\bibfield  {journal}
  {\bibinfo  {journal} {Nature Photonics}\ }\textbf {\bibinfo {volume} {7}},\
  \bibinfo {pages} {210} (\bibinfo {year} {2013})},\ \Eprint
  {http://arxiv.org/abs/1209.5774} {arXiv:1209.5774} \BibitemShut {NoStop}%
\bibitem [{\citenamefont {Klyshko}(1980)}]{Klyshko80}%
  \BibitemOpen
  \bibfield  {author} {\bibinfo {author} {\bibfnamefont {D.~N.}\ \bibnamefont
  {Klyshko}},\ }\href@noop {} {\bibfield  {journal} {\bibinfo  {journal} {Sov.
  J. Quantum Electron.}\ }\textbf {\bibinfo {volume} {10}},\ \bibinfo {pages}
  {1112} (\bibinfo {year} {1980})}\BibitemShut {NoStop}%
\bibitem [{onl()}]{onlinedata}%
  \BibitemOpen
  \href@noop {} {}\bibinfo {howpublished} {Raw data from each experimental run
  will be made available online.}\BibitemShut {Stop}%
\bibitem [{\citenamefont {Bierhorst}(2013)}]{Bierhorst2013}%
  \BibitemOpen
  \bibfield  {author} {\bibinfo {author} {\bibfnamefont {P.}~\bibnamefont
  {Bierhorst}},\ }\href@noop {} {\  (\bibinfo {year} {2013})},\ \Eprint
  {http://arxiv.org/abs/1312.2999} {arXiv:1312.2999} \BibitemShut {NoStop}%
\bibitem [{\citenamefont {Zhang}\ \emph {et~al.}(2011)\citenamefont {Zhang},
  \citenamefont {Glancy},\ and\ \citenamefont {Knill}}]{Zhang2011}%
  \BibitemOpen
  \bibfield  {author} {\bibinfo {author} {\bibfnamefont {Y.}~\bibnamefont
  {Zhang}}, \bibinfo {author} {\bibfnamefont {S.}~\bibnamefont {Glancy}}, \
  and\ \bibinfo {author} {\bibfnamefont {E.}~\bibnamefont {Knill}},\ }\href
  {\doibase 10.1103/PhysRevA.84.062118} {\bibfield  {journal} {\bibinfo
  {journal} {Phys. Rev. A}\ }\textbf {\bibinfo {volume} {84}},\ \bibinfo
  {pages} {062118} (\bibinfo {year} {2011})},\ \Eprint
  {http://arxiv.org/abs/1108.2468} {arXiv:1108.2468} \BibitemShut {NoStop}%
\bibitem [{\citenamefont {Zhang}\ \emph {et~al.}(2013)\citenamefont {Zhang},
  \citenamefont {Glancy},\ and\ \citenamefont {Knill}}]{Zhang2013}%
  \BibitemOpen
  \bibfield  {author} {\bibinfo {author} {\bibfnamefont {Y.}~\bibnamefont
  {Zhang}}, \bibinfo {author} {\bibfnamefont {S.}~\bibnamefont {Glancy}}, \
  and\ \bibinfo {author} {\bibfnamefont {E.}~\bibnamefont {Knill}},\ }\href
  {\doibase 10.1103/PhysRevA.88.052119} {\bibfield  {journal} {\bibinfo
  {journal} {Phys. Rev. A}\ }\textbf {\bibinfo {volume} {88}},\ \bibinfo
  {pages} {052119} (\bibinfo {year} {2013})},\ \Eprint
  {http://arxiv.org/abs/1303.7464} {arXiv:1303.7464} \BibitemShut {NoStop}%
\bibitem [{\citenamefont {Knill}\ \emph {et~al.}(2015)\citenamefont {Knill},
  \citenamefont {Glancy}, \citenamefont {Nam}, \citenamefont {Coakley},\ and\
  \citenamefont {Zhang}}]{knill2015}%
  \BibitemOpen
  \bibfield  {author} {\bibinfo {author} {\bibfnamefont {E.}~\bibnamefont
  {Knill}}, \bibinfo {author} {\bibfnamefont {S.}~\bibnamefont {Glancy}},
  \bibinfo {author} {\bibfnamefont {S.~W.}\ \bibnamefont {Nam}}, \bibinfo
  {author} {\bibfnamefont {K.}~\bibnamefont {Coakley}}, \ and\ \bibinfo
  {author} {\bibfnamefont {Y.}~\bibnamefont {Zhang}},\ }\href {\doibase
  10.1103/PhysRevA.91.032105} {\bibfield  {journal} {\bibinfo  {journal} {Phys.
  Rev. A}\ }\textbf {\bibinfo {volume} {91}},\ \bibinfo {pages} {032105}
  (\bibinfo {year} {2015})}\BibitemShut {NoStop}%
\end{thebibliography}%


%

\end{document}